\documentclass[iop]{emulateapj}
\usepackage{color}
\usepackage{graphicx}
\usepackage{longtable}
\usepackage{draftcopy}
\usepackage{threeparttable}
\usepackage{multirow}
\slugcomment{version \today: fm}
\shorttitle{}
\shortauthors{}

\begin{document}
\title{The Two-Dimensional Spatial Distributions of the Globular Clusters and Low-Mass X-ray Binaries of NGC4649.}
\author{R.~D'Abrusco\altaffilmark{1}, G.~Fabbiano\altaffilmark{1}, S.~Mineo\altaffilmark{2,1}, 
J.~Strader\altaffilmark{3}, T.~Fragos\altaffilmark{1}, D.-W.~Kim\altaffilmark{1}, B.~Luo\altaffilmark{4} \& 
A.~Zezas\altaffilmark{1,5,6}}

\altaffiltext{1}{Harvard-Smithsonian Astrophysical Observatory, 60 Garden Street, Cambridge, MA 02138, USA}
\altaffiltext{2}{Department of Physics, University of Durham, South Road, Durham DH1 3LE, UK}
\altaffiltext{3}{Department of Astronomy, Michigan State University, 567 Wilson Road, East Lansing, 
MI  48824-2320, USA}
\altaffiltext{4}{Department of Astronomy \& Astrophysics, 525 Davey Lab, The Pennsylvania State University, 
University Park, PA 16802, USA}
\altaffiltext{5}{Physics Department and Institute of Theoretical and Computational Physics, 
University of Crete, 71003 Heraklion, Crete, Greece}
\altaffiltext{6}{Foundation of Research and Technology, 71003, Heraklion, Crete, Greece}
\begin{abstract}

We report significant anisotropies in the projected two-dimensional (2D) spatial distributions of Globular 
Clusters (GCs) of the giant Virgo elliptical galaxy NGC4649 (M60). Similar features are found in the 
2D distribution of low-mass X-ray binaries (LMXBs), both associated with GCs and in the stellar field. 
Deviations from azimuthal 
symmetry suggest an arc-like excess of GCs extending north at 4-15 kpc galactocentric radii in the eastern 
side of major axis of NGC4649. This feature is more prominent for red GCs, but still persists in the 2D distribution 
of blue GCs. High and low luminosity GCs also show 
some segregation along this arc, with high-luminosity GCs preferentially located in the southern end 
and low-luminosity GCs in the northern section of the arc. GC-LMXBs follow the anisotropy of 
red-GCs, where most of them reside; however, a significant 
overdensity of (high-luminosity) field LMXBs is present to the south of the GC arc. These results suggest 
that NGC4649 has experienced mergers and/or multiple accretions of less massive 
satellite galaxies during its evolution, 
of which the GCs in the arc may be the fossil remnant. We speculate that the observed anisotropy in the 
field LMXB spatial distribution indicates that these X-ray 
binaries may be the remnants of a star formation event connected with the merger, or maybe be ejected 
from the parent red GCs, if the bulk motion of these clusters is significantly affected by dynamical friction.
We also detect a luminosity enhancement in the X-ray source population of the 
companion spiral galaxy NGC4647. We suggest that these may be younger high mass X-ray binaries 
formed as a result of the tidal interaction of this galaxy with NGC4649.

\end{abstract}


\section{Introduction}
\label{sec:intro}

Recent work is bringing forth a picture of complex and diverse Globular Cluster (GC) populations 
in elliptical galaxies, consistent with 
several merging episodes of which the different GC populations are the fossil 
remnant~(\citealt{strader2011}, for M87;~\citealt{blom2012}, for NGC4365). Our work on the 
elliptical galaxy NGC4261 has shown large-scale spiral-like features in the two-dimensional (2D) 
GC distribution~\citep{bonfini2012,dabrusco2013}, which would have been unreported in the study 
of radial distributions. This continuing merging evolution would be 
expected in the framework of simulations based on the $\Lambda$CDM model of hierarchical galaxy 
formation~\citep{white1978,dimatteo2005}. 

The $\Lambda$CDM model of hierarchical galaxy formation~\citep{white1978} has been successful in 
reproducing several results of both deep and large sky surveys 
(e.g.,~\citealt{navarro1994},~\citealt{dimatteo2005},~\citealt{navarro2010}). However, direct observational 
validation of continuous galaxy evolution via satellite merging from observations of galaxies in the local 
universe is only recently appearing in the literature, and has been mostly limited to the Milky Way and 
Local Group. Large-scale surveys of the stellar population of the Milky Way have shown an increasing complex 
dynamical environment of satellite dwarf galaxies interacting gravitationally with and being accreted by our galaxy 
(e.g. the Sloan Digital Sky Survey results of~\citealt{belokurov2006}). The Sagittarius Stream~\citep{ibata1994} 
is a well-studied example of this phenomenon. A wide-area survey of the M31-M33 region confirms this picture, 
detecting streams and dwarf galaxies in the halo of M31 and in the region between M31 and 
M33~\citep{mcconnachie2009}.

Although these subtle details may not be perceivable beyond our Local Group, GCs may provide a marker 
of these interactions in farther away systems. Observationally, for example, the Sagittarius dwarf/stream is 
associated with several GCs (e.g.,~\citealt{salinas2012},~\citealt{forbes2010}). Although the parent 
dwarf galaxy may be disrupted, and therefore hard to be detected against the background of the more 
luminous dominant galaxy, GC may retain the information of the encounter. This is suggested by 
simulations - albeit limited to the tidal evolution of GCs in dwarf spheroidal dark matter 
halos - which show that GCs with combined high mass and high density will survive the 
encounter~\citep{penarrubia2009}. 

The giant elliptical galaxy NGC4649 (M60) in the Virgo cluster was recently surveyed with several 
Chandra and HST exposures, completely covering the area within the D25 ellipse~\citep{devaucouleurs1991}
and extending with variable coverage to larger radii. The resulting catalogs of GCs~\citep{strader2012}
and Low Mass X-ray Binaries (LMXBs)~\citep{luo2013} give us a unique opportunity to study 
the properties of these populations and to explore further the GC-LMXB 
connection~\citep[see review,][]{fabbiano2006}, 
with full spatial coverage. NGC4649 is a relatively isolated galaxy, except for the close neighbor spiral 
NGC4647, with which NGC4649 may be tidally interacting~\citep{lanz2013}.

In the first follow-up paper based on these Chandra and HST data~\citep{mineo2013}, we studied 
the radial distributions of the GCs and LMXBs populations and compared them with that of the 
optical surface brightness, 
which is a good proxy of the stellar mass for the old stellar population of NGC4649. The radial 
distributions of red and blue GCs differ, as generally observed in other 
galaxies~\citep[see review,][]{brodie2006,strader2012}. The blue GC radial profile is 
definitely wider than that of the stellar light. The red GCs profile is steeper, close to that of the stellar 
light, with a noticeable ``dip'' in the denser centermost region (r$<\!40^{\prime\prime}\!\sim$ 3 kpc at the 
NGC4649 distance of 16.5 Mpc~\citep{blakeslee2009}), 
where tidal disruption of GCs may be more efficient. The LMXBs associated with 
GCs follow the same radial distributions as their parent GC populations. Roughly three times more of 
these LMXBs are found in red rather than in blue GCs, as generally observed in early-type 
galaxies~(\citealt{kundu2002};~\citealt{kim2009};~\citealt{mineo2013}; see~\cite{ivanova2012} for a 
theoretical explanation).
The field LMXBs - those without a GC counterpart - 
follow the stellar 
light, i.e., the mass distribution of old stars, within the 
$D_{25}$ ellipse~\citep{kim2004,gilfanov2004}.~\cite{mineo2013} also report a departure of the LMXB 
radial distribution 
from that of the stellar mass at larger galactocentric radii, which could also be associated with more 
luminous sources. As discussed there, this over-luminosity may 
point to a younger field LMXB population: evolutionary models of native field LMXB populations 
have shown that LMXBs 
become fainter with the increasing age of the parent stellar population~\citep{fragos2013a,fragos2013b}.

Here we complete our study of the spatial distribution of the GC populations in NGC4649, by 
examining their 2D observed distributions, to explore the presence of irregularities that 
may point to mergers or satellite accretion by NGC4649. Given the association and possible 
evolutionary relation between GCs and LMXBs~\citep[see][and refs. therein]{fabbiano2006}, 
we also extend this study to the LMXB population of NGC4649. In particular we seek to explore if the 
regularity observed in the radial distributions of these sources still persists in the azimuthal dimension. 
 
In Section 2 we describe the catalogs of GCs and LMXBs used in this paper. These data were analyzed 
with the method described in~\cite{dabrusco2013}. This method is based on the K-Nearest 
Neighbor (KNN) density estimator~\citep{dressler1980}, augmented by Monte Carlo simulations, to 
estimate the statistical significance of any spatial feature (Section~\ref{sec:method}). The results are 
presented in Section~\ref{sec:densityresidual} and discussed in Section~\ref{sec:discussion}. We 
summarize our finding in Section~\ref{sec:conclusions}. In this paper we adopt a distance of 
16.5 Mpc to NGC4649~\citep{blakeslee2009}. At this distance, an angular separation of 1$^{\prime}$ 
corresponds to a linear distance of $\sim$4.8 kpc.

\section{Data: GCs and LMXBs}
\label{sec:data}

The GC and LMXB samples used in this work are the results of the joint Chandra-HST 
large-area mapping of NGC4649 (PI Fabbiano\footnote{Public {\it Chandra} data can be retrieved 
through the CXC, at the following URL: http://cxc.harvard.edu/. HST public data can be searched 
and retrieved using the MAST archive, available at the URL: http://archive.stsci.edu/.}). The GC
sample is based on the GC catalog
extracted by~\cite{strader2012} from six HST ACS pointings of NGC4649. It contains
1603 GCs with photometry in $g$ and $z$ filters, selected by requiring 
$z\!<\!24$ and $0.5\!<g-z\!<\!2.0$ and discarding all sources with size consistent with the PSF 
(presumed to be foreground stars). This catalog includes 841 red GCs with 
$g\!-\!z\!>\!1.18$, and 762 blue GCs with $g\!-\!z\!\leq\!1.18$. The color threshold 
was derived from~\cite{strader2012} after verifying that significant color substructures 
are visible in the whole interval of galactocentric distances explored by the catalog of GCs.
While the shapes and positions of the red and blue features in the color distribution
of GCs (see Figure~6 in~\cite{strader2012}) show a slight dependence on the galactocentric 
distance within NGC4649, the $g\!-\!z\!=\!1.18$ value allows a clear separation of 
the two GCs color classes over the whole range of distances. For our analysis we have also 
considered a split of the sample in luminosity, resulting in 467 high-luminosity GCs with $g\!\leq\!23$, 
and 1136 low-luminosity GCs with $g\!>\!23$. 
As discussed later in Section~\ref{sec:method}, small variations of both color and 
luminosity boundaries do not alter the results of our analysis. In this particular case, 
the magnitude threshold used to define low- and high-luminosity GCs was chosen only 
to provide luminosity classes of similar sizes (see Table~\ref{tab:summary}).

\begin{figure*}[]   
	\includegraphics[height=8cm,width=8cm,angle=0]{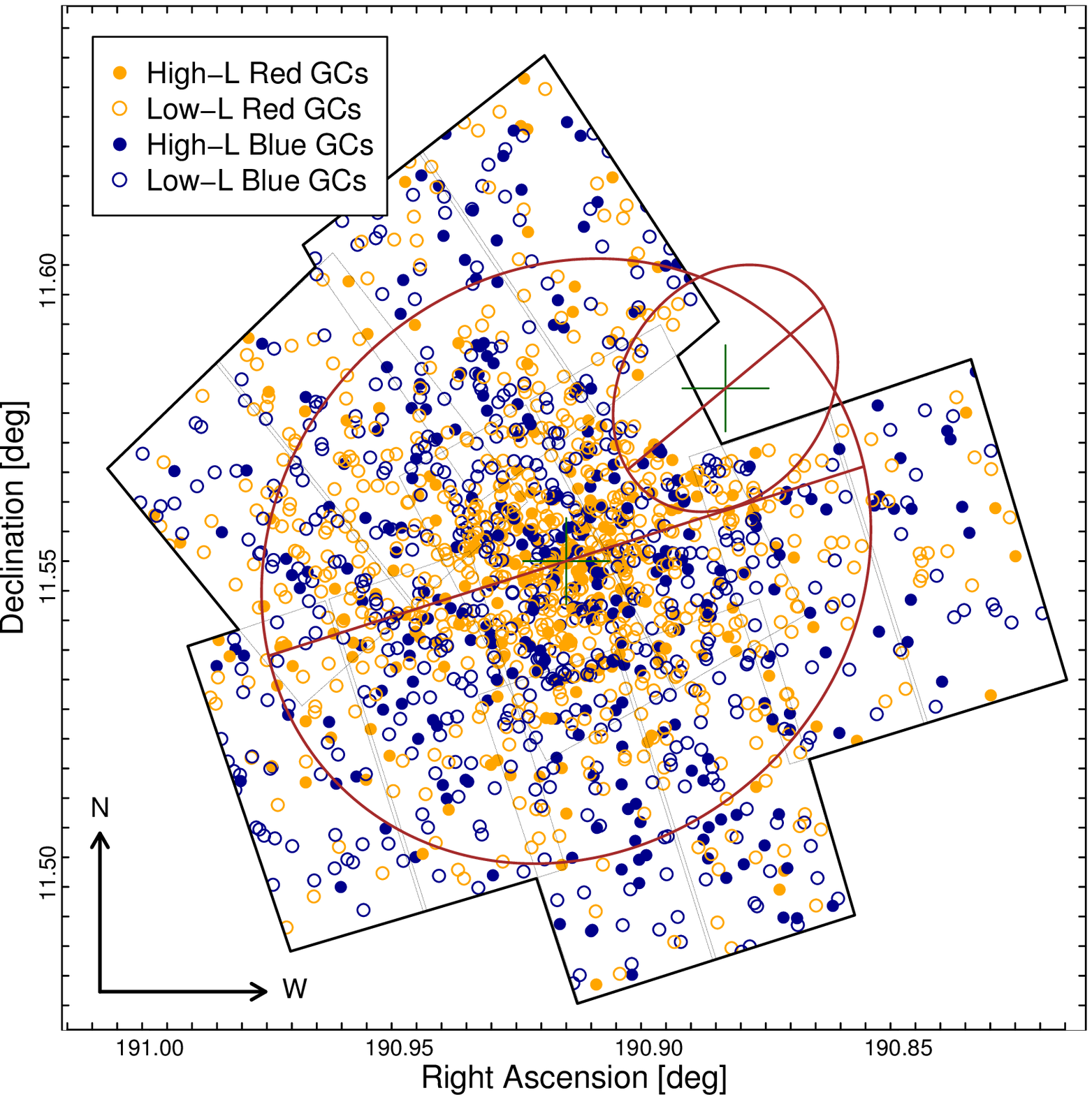}
	\includegraphics[height=8cm,width=8cm,angle=0]{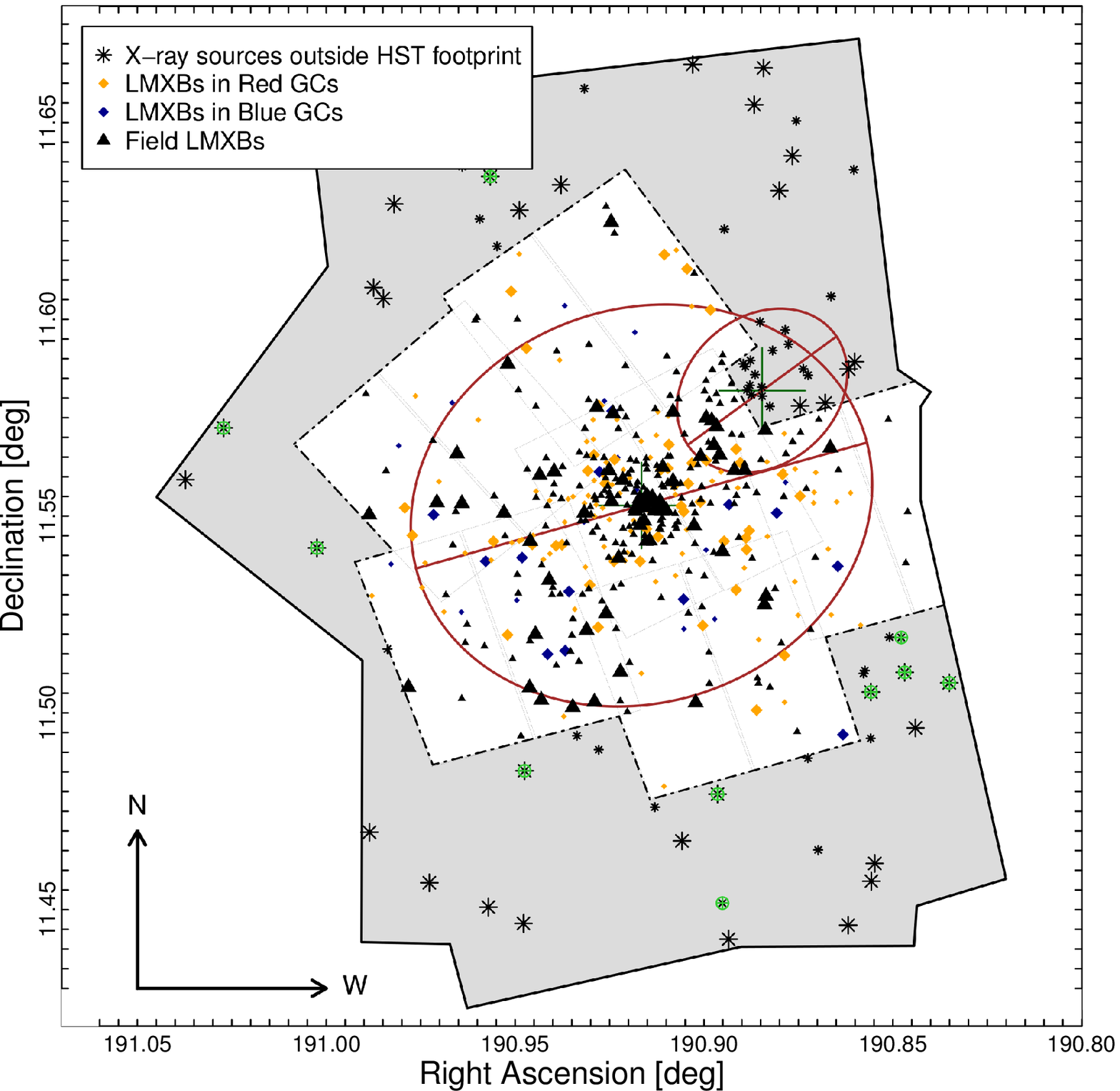}	
	\caption{Left: GC positions in NGC4649. Blue and orange symbols identify blue and red 
	GCs, respectively. Solid and open symbols identify high- and low-luminosity GCs.
	The large and small ellipses are the $D_{25}$ isophotes of NGC4649 and the neighboring 
	spiral galaxy NGC4647 from~\cite{devaucouleurs1991}. The squares identify the 
	footprint of the HST observations used by~\cite{strader2012} to extract the catalog
	of GCs. 
	Right: LMXBs in NGC4649. GC-LMXBs in blue and red GCs 
	are indicated with blue and orange symbols, while field LMXBs are plotted as black triangles.
	High- and low-luminosity LMXBs are shown as large and small symbols respectively.
	The outer polygon shows the boundaries of the {\it Chandra} footprint. The enclosed white area shows the HST 
	footprint used to extract the GCs sample. The large and small ellipses are 
	the $D_{25}$ isophotes of NGC4649 and NGC4647, respectively. The major axis of NGC4649  
	is also shown~\citep{devaucouleurs1991}.
	The green stars are the LMXBs associated to GCs from the ground-based catalog of~\cite{lee2008}.}
	\label{fig:positions}
\end{figure*}

Figure~\ref{fig:positions} (left) shows the spatial distributions of the GCs in the plane of the sky. 
The HST pointings overlap marginally with the $D_{25}$ elliptical 
isophote of the neighboring spiral galaxy NGC4647. We have excluded all the GCs within the 
NGC4647 $D_{25}$ ellipse from the sample used to evaluate the observed density profile of GCs in 
NGC4649 (see Section~\ref{sec:method}), but we have kept them in the density maps. 

The LMXB sample is from the catalog of~\cite{luo2013}, where identifications with GCs 
are based on~\cite{strader2012}. This catalog was derived from six {\it Chandra} 
ACIS-S3 pointings of NGC4649, reaching a total exposure of 299 ks and yielding a total
of 501 X-ray sources with 0.3-8.0 keV luminosities ranging from 
$9.3\!\times\!10^{36}$ to $5.4\!\times\!10^{39}$ ergs s$^{-1}$. Out of the total 427 LMXBs 
within the HST combined footprint, 161 are GC-LMXBs and 266 are found in the stellar field 
of NGC4649 (field LMXBs). Moreover, 74 LMXBs were
detected in the region external to the HST footprint (see~\citealt{mineo2013}). 
Of the X-ray sources detected in the area not overlapped by the HST observations used 
to extract the catalog of GCs used in this paper, 
twelve of them can be identified with GCs in the ground-based catalog 
of~\cite{lee2008}; the other X-ray sources in this area are likely to be mostly background Active 
Galactic Nuclei (AGNs), based on the contamination estimates of~\cite{luo2013}. 
To avoid the large 
uncertainties on the sample of X-ray sources detected at larger radii, in this paper we will derive the 
density reconstruction of only the GCs and field LMXBs located within the HST footprint.
Only $\sim\!20$ background AGNs are expected within the $D_{25}$ elliptical isophote of NGC4649; 
these are likely to contaminate the field LMXB sample, because the GC identification excludes this contamination 
in the sample of GC-LMXBs. We have also explored luminosity classes
based on a threshold of $L_{\mathrm{X}}\!=\!10^{38}$ erg 
s$^{-1}$ for the LMXBs. The two luminosity classes comprise 339 
and 162 sources respectively. The higher luminosity class may contain a larger proportion of black 
hole binaries or of younger LMXBs~(\citealt{fragos2013a,tzanavaris2013}). The spatial distribution of 
the LMXBs is shown in Figure~\ref{fig:positions} (right). The number of members of each class of 
sources investigated in this paper is reported in Table~\ref{tab:summary}.

\begin{table*}[h]
	\centering
	\caption{Samples of GCs and LMXBs.}
	\begin{tabular}{lccccc}
	\tableline
			& $N_{\mathrm{tot}}$	&$N_{\mathrm{red}}$ &$N_{\mathrm{blue}}$ 	& $N_{\mathrm{HighL}}$ & $N_{\mathrm{LowL}}$	\\			
	GCs\footnote{ In square brackets, the number of low-luminosity and high-luminosity GCs for red and blue classes, respectively.}		& 1603				&841[642/198]	  	& 762[493/269]			& 467			        &  1136				\\
	\tableline
	\tableline
			& $N_{\mathrm{tot}}$	&$N_{\mathrm{GCs}}$ &$N_{\mathrm{field}}$ 	& $N_{\mathrm{HighL}}$ & $N_{\mathrm{LowL}}$	\\			
	LMXBs\footnote{In square brackets, the number of LMXBs associated to blue and red GCs, respectively. 
	In parenthesis, the number of X-ray sources located outside the HST footprint.}	& 501 	& 161[28/133]	& 266(74)	& 162	&  339	\\
	\tableline
	\end{tabular}\\
	\label{tab:summary}
\end{table*}

\begin{figure*}[]
	\includegraphics[height=8cm,width=8cm,angle=0]{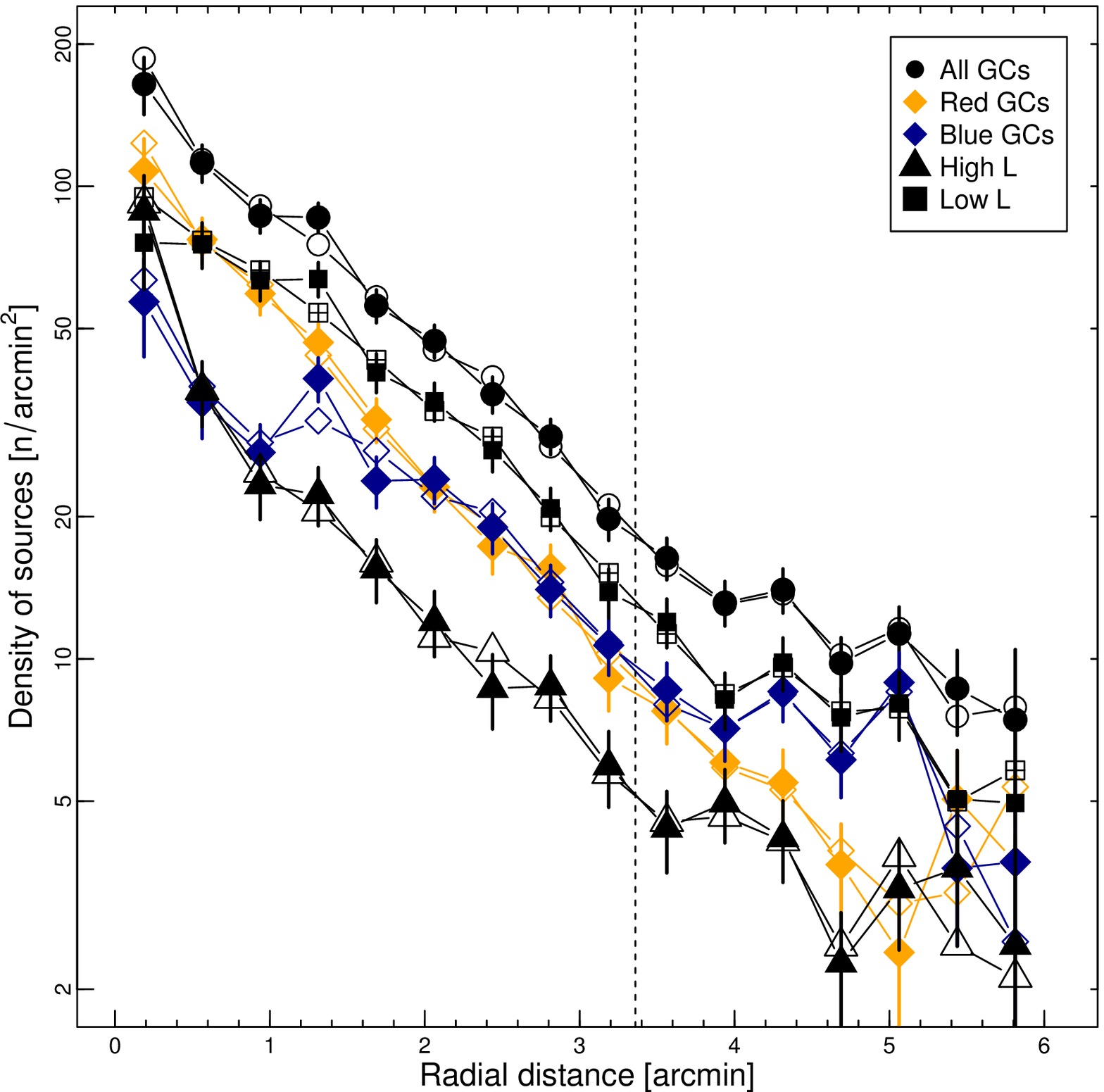}
	\includegraphics[height=8cm,width=8cm,angle=0]{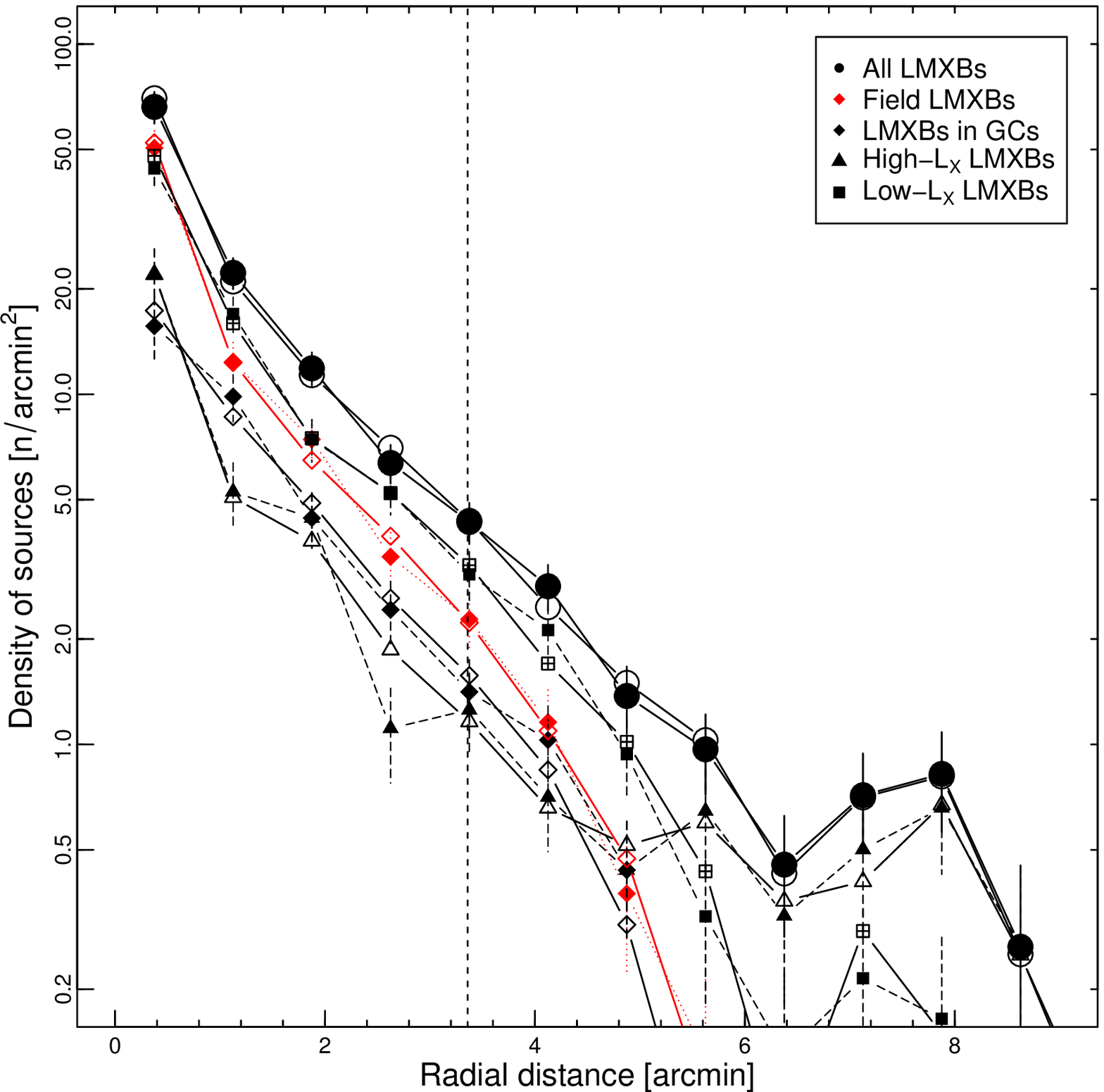}	
	\caption{Left: Observed (solid) and simulated 
	(open) GC radial density profiles integrated in elliptical annuli. All GCs: black circle, red GCs: orange
	diamonds, blue GCs: blue diamonds, high-luminosity GCs: black triangles, and low-luminosity: black squares. 
	The vertical dotted line marks the maximum radial distance of the $D_{25}$ ellipse.
	Right: LMXB observed (solid) and simulated 
	(open) radial density profiles integrated in elliptical annuli. All LMXBs (black circles),  
	field LMXBs (red diamonds), GC LMXBs (black diamonds), high-luminosity LMXBS (black triangles) 
	and low-luminosity LMXBs (black squares) are shown. The vertical dotted 
	line marks the maximum radial distance of the $D_{25}$ ellipse of NGC4649.}
	\label{fig:radialprofiles}
\end{figure*}

\section{Method}
\label{sec:method}

We determined the 2D spatial distributions of GCs and LMXBs by applying a method based on 
the KNN density estimator~\citep{dressler1980}. We used simulations
to evaluate the significance of the results. A detailed explanation of our application of the KNN 
method can be found in~\cite{dabrusco2013}. 

The assumption behind the KNN method is that the density is locally constant.
The uncertainty on 
the KNN density scales with the square root of $K$, where $K$ is the index of the nearest neighbor 
used to calculate the density and the fractional accuracy of the method increases 
with increasing K at the expense of the spatial resolution. For each sample discussed in this paper, 
we calculated the 2D surface densities for $K$ ranging from 2 to 9, and a regular grid with 
spacing $\Delta\!(\mathrm{R.A})$ $\sim\!0.005^{\circ}$ ($\sim18^{\prime\prime}$) and 
$\Delta\!(\mathrm{Dec})$ $\sim\!0.004^{\circ}$ ($\sim\!14^{\prime\prime}$) for the GCs, 
and  $\Delta\!(\mathrm{R.A})$ 
$\sim\!0.006^{\circ}$ ($\sim\!22^{\prime\prime}$) and $\Delta\!(\mathrm{Dec})$ $\sim\!0.006^{\circ}$ 
($\sim\!22^{\prime\prime}$) for the LMXBs. These grids define 2D cells of ``pixels''. This choice of 
grids is justified by the different areas covered by 
the HST and {\it Chandra} observations and produces similar average number of source per pixel
for both GCs and LMXBs. This choice permits to compare meaningfully the statistical reliability of 
the features determined in density and residual maps for both classes of sources.
The density in the boundary pixels was weighted according to 
the fraction of the pixel within the observed region. 

To assess the statistical significance of features suggested by the KNN density maps 
(hereafter ``observed'' density maps), we employed Monte-Carlo simulations, 
using the observed radial distributions of each of the classes of GCs and LMXBs investigated, 
as seeds. The azimuthal distributions 
were independently extracted from a uniformly random distribution 
between 0$^{\circ}$ and $360^{\circ}$. 
A geometrical correction on the expected number of GCs per
elliptical annulus was applied to take into account the eccentricity of the galaxy. 
We performed 
$N_{\mathrm{Sim}}\!=\!1000$ simulations for each of the samples described in 
Table~\ref{tab:summary}, and analyzed the results as done with the observed data (see above). 
Figure~\ref{fig:radialprofiles} compares observed and simulated mean radial density profiles for 
the GC and LMXB samples, respectively.

To characterize the deviations from the average 2D trends of the GC and LMXB distributions, 
we calculated maps of residuals by subtracting pixel-by-pixel the mean simulated density map 
from the corresponding observed density map, and normalizing to the
value of the density in the average simulated map. 

%The residual $R_{i}$ 
%of the $i$-th pixel of the map is thus defined as:
%
%\begin{equation}
%	R_{i}\!=\frac{\!(O_{i}\!-\!<\!S\!>_{i})}{<\!S\!>_{i}}
%	\label{eq:residuals}
%\end{equation}

The pixel-by-pixel distributions of the simulated KNN densities are well approximated by 
Gaussians, simplifying the calculation of the statistical significance of the residuals.
To evaluate the latter, for each set of simulated density maps we calculated the fraction of pixels 
with values above the 90-th percentile of the densities in the observed maps (the ``extreme'' pixels). 
Since these extreme pixels in both observed density and residual maps tend to be 
spatially correlated, we also evaluated the fraction of simulations with at least one group of contiguous 
extreme pixels (with area equal to 18 pixels) as large as the observed. 
Since we did not impose any specific geometry to the groups of simulated contiguous
extreme pixels, these fractions are upper limits to the fraction expected for a given spatial distribution 
of residuals. The results are given in Section~\ref{sec:densityresidual}.

As discussed above, the radial distributions of the simulated distributions of sources follow the 
observed radial distributions, while the azimuthal distribution is uniform in the $[0,\!2\pi]$ 
interval. If the incompleteness of the observed distribution of sources is a function only of the radial 
distance from the center without azimuthal dependencies, it will affect 
in the same way both observed and simulated density maps. Accordingly, in this paper 
we have not included completeness
corrections for either GC and LMXB samples because both populations are affected by incompleteness 
which only depends on the radial dependence from the center of the NGC4649 galaxy 
(see~\citealt{mineo2013,luo2013} for GCs and LMXBs incompleteness maps respectively).

\begin{table*}
	\centering
	\caption{Fractions of simulated density maps of the population of GCs and LMXBs
	with the number of extreme pixels (i.e., pixels with density values exceeding the 90-th 
	percentile of the observed pixel density 
	distribution) larger than the number of observed extreme pixels. Values in parenthesis 
	refer to the fraction of simulated density maps with at least one 
	group of contiguous extreme pixels as large as the groups of contiguous extreme pixels
	in the observed density maps (see details in Section~\ref{sec:method}).
	These fractions were determined by counting the number of simulated density
	maps with at least one group of contiguous extreme pixels equal or larger than
	the group of contiguous extreme pixels observed over-density regions.}
	\begin{tabular}{lccccccc}
	\tableline
	Density				&				&			&			&			&			&				\\
		 				&$K\!=\!4$		&$K\!=\!5$	&$K\!=\!6$	&$K\!=\!7$	&$K\!=\!8$	&$K\!=\!9$		\\
	\tableline
	All GCs (red$+$blue)	&99.5\%(1.8\%)		&49.5\%(1.2\%)	&6\%(0.3\%)	&0\%(0\%)	&0\%(0\%)	&0\%(0\%)		\\
	Red GCs				&96\%(3.1\%)		&25.9\%(2.5\%)	&2.3\%(1.4\%)	&0\%(0.3\%)	&0\%(0\%)	&0\%(0\%)		\\
	Blue GCs				&100\%(4.7\%)		&60.5\%(1.2\%)	&4\%(0\%)	&1.2\%(0\%)	&0\%(0\%)	&0\%(0\%)		\\	
	High-L GCs			&100\%(3.4\%)		&74\%(0.9\%)	&15\%(0.2\%)	&2.8\%(0\%)	&0\%(0\%)	&0\%(0\%)		\\	
	Low-L GCs			&99.5\%(1.3\%)		&49\%(0.3\%)	&5.7\%(0\%)	&0\%(0\%)	&0\%(0\%)	&0\%(0\%)		\\	
	\tableline
	All LMXBs				&97\%(73\%)		&86\%(56.4\%)	&68\%(34.8\%)		&45.5\%(12.9\%)	&33.5\%(8\%)	&21.5\%(0.5\%)		\\
	GC-LMXBs		&85.5\%(54.6\%)	&64.5\%(33.1\%)&45.5\%(12.2\%)	&33.5\%(3.5\%)		&25.5\%(1.4\%)	&20.5\%(0.3\%)		\\
	Field LMXBs			&85\%(34\%)		&64\%(28.8\%)	&52\%(17.5\%)	  	&43\%(11.8\%)		&12\%(5.4\%)	&20\%(0\%)		\\	
	High-L LMXBs 			&99.5\%(87.4\%)	&94\%(76.9\%)	&88\%(65.3\%)		&71\%(43.9\%)		&60\%(19.3\%)	&53.5\%(0.8\%)		\\	
	Low-L LMXBs 			&61\%(17.9\%)		&28\%(12.8\%)	&15.5\%(3.7\%)		&7.5\%(0.5\%)		&4\%(0\%)	&3\%(0\%)		\\	
	\tableline	
	Residuals				&				&			&			&			&			&			\\
		 				&$K\!=\!4$		&$K\!=\!5$	&$K\!=\!6$	&$K\!=\!7$	&$K\!=\!8$	&$K\!=\!9$	\\
	\tableline
	All GCs (red$+$blue)	&65.2\%(10.5\%)	&8.2\%(4.8\%)	&1.1\%(0.7\%)	&0\%(0\%)	&0\%(0\%)	&0\%(0\%)	\\
	Red GCs				&54.8\%(8.9\%)		&4.2\%(1.1\%)	&0.8\%(0.5\%)	&0\%(0.1\%)	&0\%(0\%)	&0\%(0\%)	\\
	Blue GCs				&71.8\%(0.2\%)		&9.1\%(0\%)	&1.7\%(0\%)	&0\%(0.4\%)	&0\%(0\%)	&0\%(0\%)	\\
	High-L GCs			&89.5\%(12.2\%)	&16\%(1.9\%)	&4.5\%(0.6\%)	&0.5\%(0\%)	&0\%(0\%)	&0\%(0\%)	\\
	Low-L GCs			&88.6\%(7.2\%)		&5.2\%(5.4\%)	&0\%(1.1\%)	&0\%(0.4\%)	&0\%(0\%)	&0\%(0\%)	\\		
	\tableline
	All LMXBs 			&78.2\%(10.5\%)	&51.7\%(4.8\%)	&22.1\%(2.7\%)		&13.5\%(0\%)	&2.3\%(0\%)	&0\%(0\%)	\\
	GC-LMXBs	&64\%(8.9\%)		&41.7\%(3.1\%)	&30.4\%(1.5\%)		&14.2\%(0\%)	&2.5\%(0\%)	&0.1\%(0\%)	\\
	Field LMXBs 			&81\%(12.2\%)		&59.4\%(9.3\%)	&43.9\%(4.7\%)		&28.7\%(2.4\%)	&6.4\%(1.3\%)	&0.5\%(0\%)	\\
	High-L LMXBs 			&85.9\%(17.2\%)	&78.4\%(12.9\%)&52.1\%(9.6\%)	&34.7\%(6.5\%)	&21\%(5\%)	&9.7\%(0.8\%)	\\
	Low-L LMXBs 			&57.4\%(7.2\%)		&43.8\%(5.4\%)	&28.5\%(3.2\%)		&13.9\%(1.6\%)	&3.2\%(0\%)	&0.1\%(0\%)	\\		
	\tableline	
	\end{tabular}
	\label{tab:statistics}
\end{table*}

\section{Results of the KNN analysis}
\label{sec:densityresidual}

For each GC and LMXB sample, Table~\ref{tab:statistics} summarizes the percentages 
of simulated maps with a number of extreme 
pixels exceeding that in the observed density and residual maps for values of $K$ ranging
from 4 to 9. Given the size of the samples
of GCs and LMXBs used in this paper, larger values of $K$, while enhancing possible low-contrast 
large-scale features, would degrade the spatial resolution of the
density and residual maps. The fraction of extreme
pixels in the simulated density and residual maps decreases 
for increasingly larger $K$ values: for GCs, it is zero
for $K\!=\!8$; for LMXBs, very small and zero values are obtained for $K\!=\!9$ in the case of 
residual maps, while significantly positive values occur even for $K\!=\!9$ for the density maps. 
However, once the spatial clustering
of the extreme pixels is taken into account, the LMXBs density maps have negligible fractions of extreme 
simulated pixels when $K\!=\!9$, making
the asymmetries at these scales unlikely to be the results of 
random statistical fluctuations. In the following we will show and discuss the 
results obtained for $K\!=\!8$ for all GCs classes and $K\!=\!9$ for LMXBs. Table~\ref{tab:numover} gives the 
number of sources of each class located within pixels associated to over-density with 
significance greater than 1, 2, and 3$\sigma$ for the GC ($K\!=\!8$) and LMXB ($K\!=\!9$) samples.
In Table~\ref{tab:numover}, the fraction of sources located in the $>\!1$, $>\!2$ and $>\!3\sigma$
over-density pixels of the map and the number of sources in excess to the expected number
(based on the radial profiles shown in Figure~\ref{fig:radialprofiles}) are shown in parentheses 
and square brackets, respectively.

\subsection{Density and Residual maps of GCs}
\label{subsec:gcdensity}

\begin{figure*}[]
	\includegraphics[height=6cm,width=6cm,angle=0]{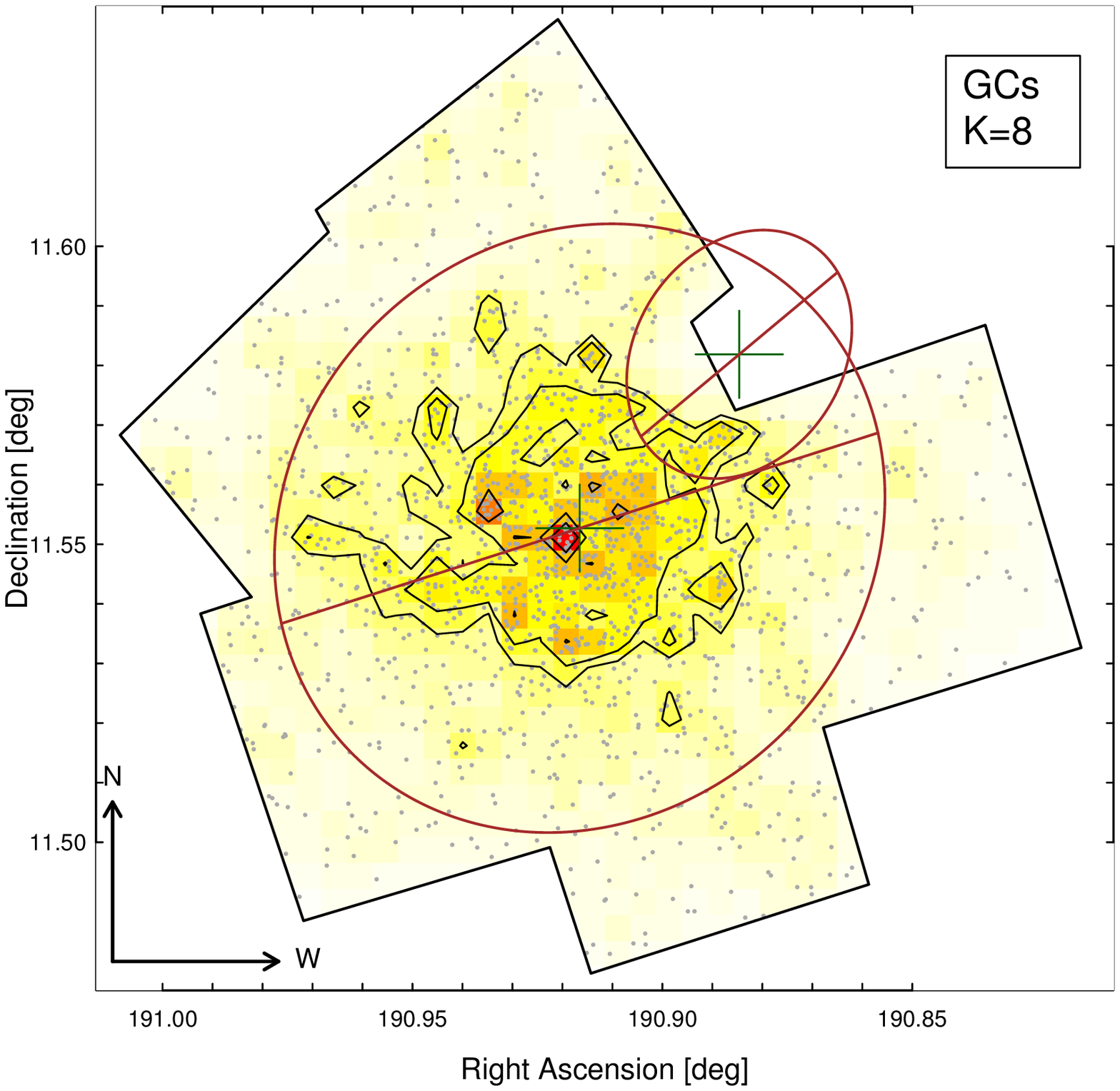}
	\includegraphics[height=6cm,width=6cm,angle=0]{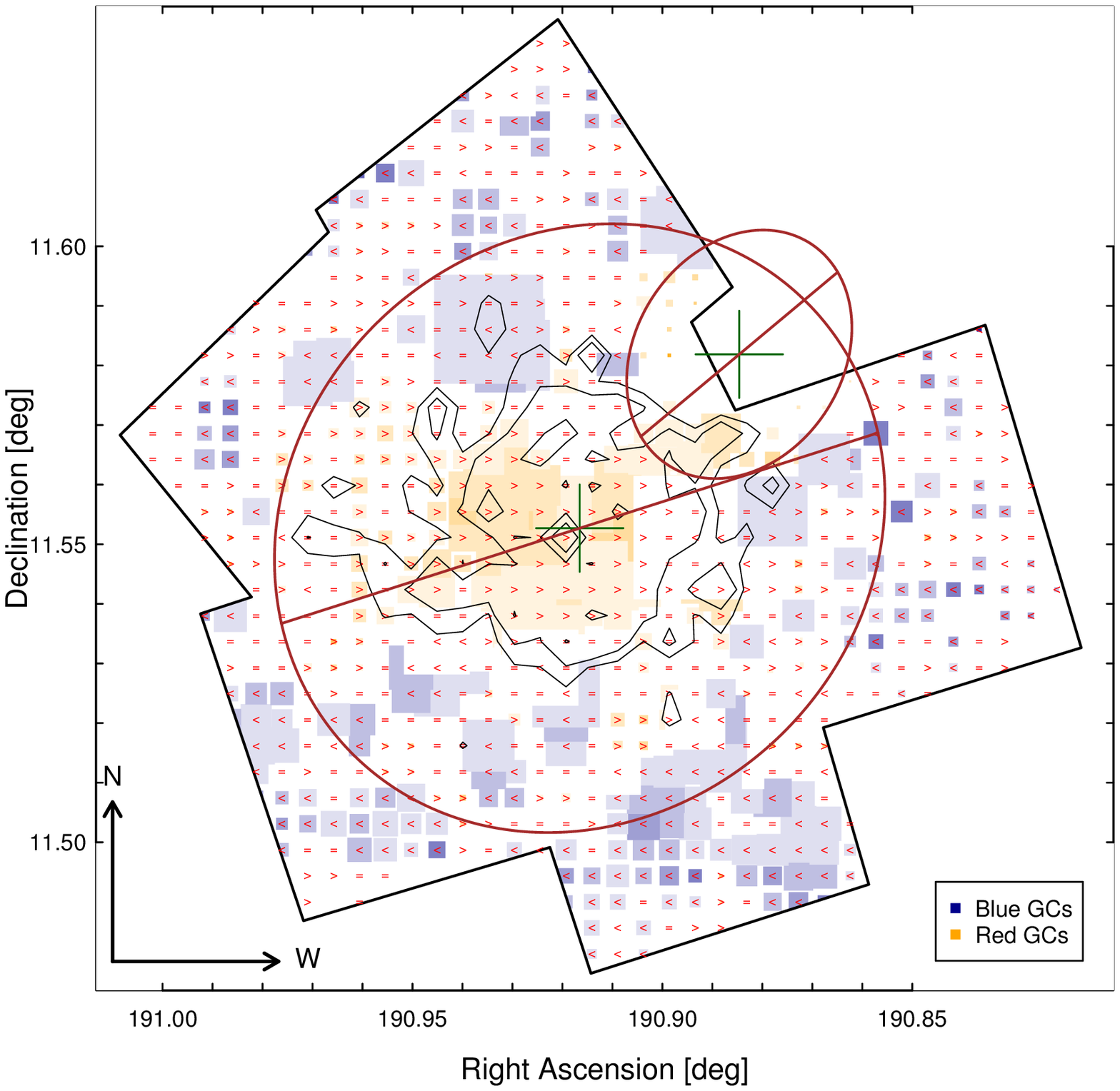}
	\includegraphics[height=6cm,width=6cm,angle=0]{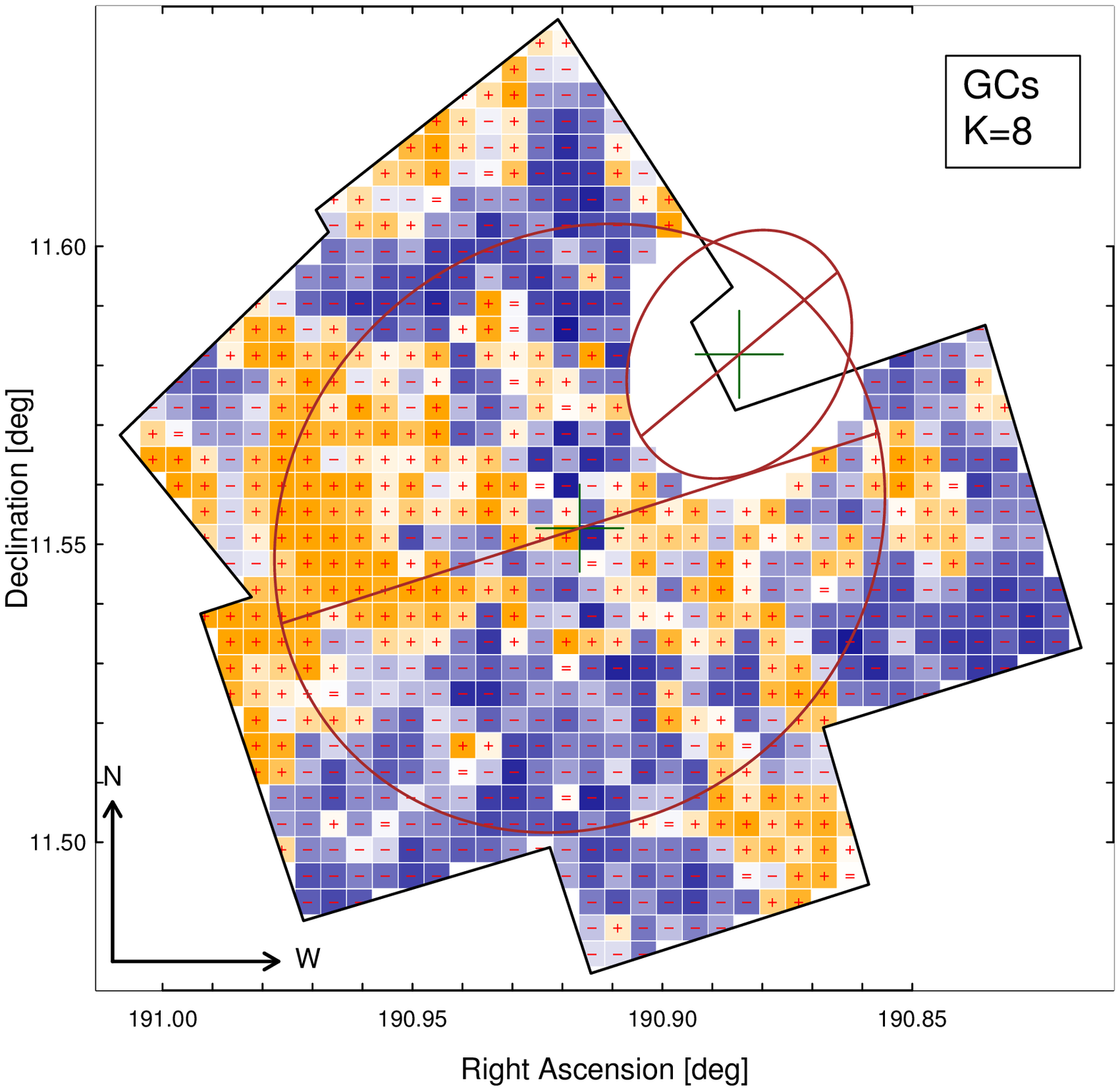}		
	\caption{Left: $K\!=\!8$ density map of the entire GC sample of NGC4649. Arbitrary isodensity contours 
	show the higher-density regions. 
	Center: Map of the average color of GCs in pixels. The size of the symbols is proportional to 
	the density of GCs distribution evaluated with the KNN and $K\!=\!8$ and the color intensity of the 
	symbols is proportional to the difference between the color threshold value $g\!-\!z\!=\!1.18$ and 
	the average color of the GCs in each pixel. Small red symbols $>$, $=$ and $<$ are drawn 
	within each pixel with average
	GCs color $g\!-\!z\!>\!1.18$, $g\!-\!z\!=\!1.18$ and $g\!-\!z\!<\!1.18$ respectively. The isodensity 
	contours reflect the density map with $K\!=\!8$ derived from the distribution of all GCs.
	Right: $K\!=\!8$ residual map of the entire GCs sample. Pixels are color-coded 
	according to the number of $\sigma$ the pixel deviates from the average. Darker colors indicate larger
	residuals: blue, negative; orange, positive. The small ``$+$'', ``$-$'' and ``$=$'' signs within each 
	pixel indicate positive, negative or null residuals respectively. The footprint of the HST observations
	used to extract the catalog of GCs and the $D_{25}$ ellipses of both NGC4649 and NGC4647 are 
	shown. The arbitrary isodensity contours in the left and center plots are only shown to highlight the 
	position of the main over-densities.}
	\label{fig:2dmapsngc4649_gc}
\end{figure*}

Figure~\ref{fig:2dmapsngc4649_gc} (left) shows the 2D KNN density map of the spatial distribution 
of the entire GC sample for $K\!=\!8$. GCs are concentrated 
in the center of the galaxy, with an over-density elongated along the major 
axis of NGC4649. Figure~\ref{fig:2dmapsngc4649_gc} (center) shows the same density map 
where pixels are color-coded according to the average color of the GCs located in each pixel. 
As expected, high-density pixels, which tend to be more 
centrally located, contain on average redder GCs. 

\begin{figure}[]
	\includegraphics[height=8cm,width=8cm,angle=0]{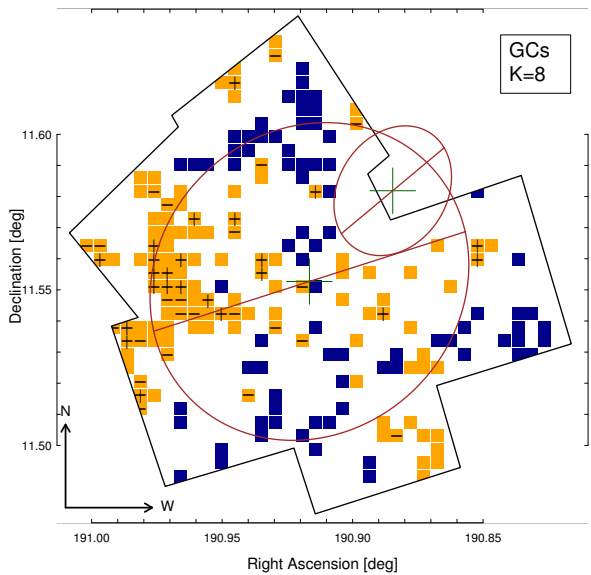}		
	\caption{Positions of the $K\!=\!8$ residuals with significance larger than 
	1$\sigma$, 2$\sigma$ and 3$\sigma$ obtained from the residual maps derived from the 
	distribution of the whole catalog of GCs in NGC4649. All the negative residuals (blue pixels)
	have significance between 1 and 2 $\sigma$. Positive (orange) pixels $\!>\!2\sigma$ are indicated 
	with ``$-$'' sign; $\!>\!3\sigma$ with a ``$+$'' sign. The alignment of the over-density 
	and under-density pixels along the major and minor axes of the galaxy, respectively, is not
	an effect of the assumption of the circularly symmetric distribution of sources in the simulations.}
        \label{fig:res2dmapsngc4649sigmas_gc}
\end{figure}

\begin{figure}[]
	\includegraphics[height=8cm,width=8cm,angle=0]{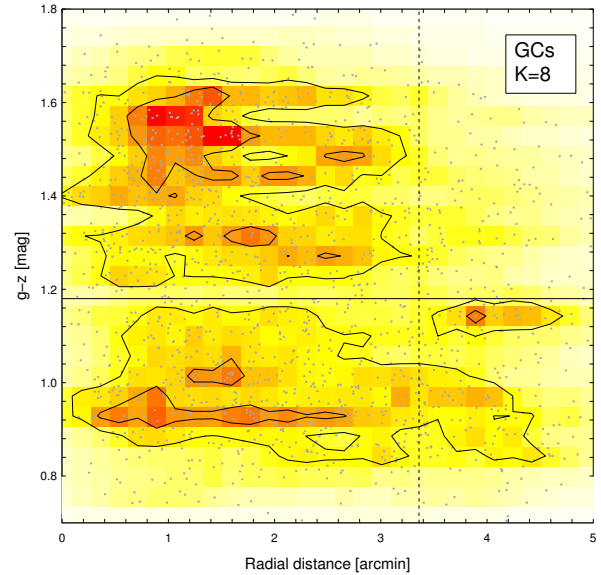}	
	\caption{Density map of GCs distribution in the plane generated by the radial distance
	and the $g\!-\!z$ color obtained with the KNN method for $K\!=\!8$. The horizontal black line shows 
	the color value used as threshold to separate red and blue GCs in this paper. The gray 
	points represent the observed positions of the GCs used to reconstruct the density. The arbitrary 
	isodensity contours are added to highlight the position of the main over-densities.}
	\label{fig:dens2dradialgmzcolor_gc}
\end{figure}

A slight excess ($\sim\!1.5\sigma$) of red GCs is found in the over-density structure observed in the 
N-E quadrant of the galaxy and along the East section of the major axis. Figure~\ref{fig:2dmapsngc4649_gc} 
(right) shows the residual map, which emphasizes 
over-densities along the major axis, extending into the N-E quadrant. 
For $K\!=\!8$, the number of all GCs located within the pixels associated to 
positive residuals is 438 ($\sim\!27\%$), 208 ($13\%$) and 127 ($\sim\!8\%$) for 1$\sigma$, 2$\sigma$ 
and 3$\sigma$ significance respectively (Table~\ref{tab:numover}). 
Figure~\ref{fig:res2dmapsngc4649sigmas_gc} shows that the majority of the pixels associated to 
positive residual values with significance larger than 2$\sigma$ and 3$\sigma$ are located in the E 
and to the N-E of the main axis of NGC4649. 

Figure~\ref{fig:dens2dradialgmzcolor_gc} shows the $K\!=\!8$ density map in the radial 
distance vs $g\!-\!z$ plane. The bi-modality of the distribution of GCs in color is clear for $r\!<\!3.4^{\prime}$
which corresponds to the major axis of the $D_{25}$ isophote, in agreement with the finding 
of~\cite{strader2012}. For $r\!>\!3.4^{\prime}$ the red, metal-rich GCs  
become less numerous, confirming that the average $g\!-\!z$ color in pixels outside the $D_{25}$ is significantly 
bluer than inner regions of the galaxy (see Figure~\ref{fig:2dmapsngc4649_gc}, center).

\begin{table*}
	\centering
	\caption{Number of GCs and LMXBs located in over-densities regions with significance larger 
	than 1 $\sigma$, 2$\sigma$ and 3$\sigma$ for all 
	classes of sources used in this paper. In parentheses, the percentage relative to the total number 
	of sources in each class, as shown in Tab.~\ref{tab:summary}, and in square brackets, the 
	number of sources in excess to the expected number.}
	\begin{tabular}{lccc}
	\tableline
				& $1\sigma$&$2\sigma$&$3\sigma$\\
	\tableline
	All GCs\footnote{Maps obtained for $K\!=\!8$.}			&438($27.3\%$)[190] &208($13\%$)[109] &127($7.9\%$)[82]\\
	Red GCs		&388($46.1\%$)[258] &184($21.9\%$)[135]&111($13.2\%$)[82]\\
	Blue GCs		&388($50.9\%$)[266] &161($21.1\%$)[117]&74($9.7\%$)[57]\\
	High-L GCs	&307($65.7\%$)[240] &109($23.3\%$)[87]&45($9.6\%$)[37]\\
	Low-L GCs	&430($37.9\%$)[243] &180($15.8\%$)[111]&82($7.2\%$)[55]\\
	\tableline
	All LMXBs\footnote{Maps obtained for $K\!=\!9$.}		&153($30.5\%$)[74] &72($14.4\%$)[40] &47($9.4\%$)[31]\\
	GC LMXBs	&60($37.3\%$)[30] &43($26.7\%$)[25]&31($19.3\%$)[20]\\
	Field LMXBs	&83($31.2\%$)[38] &39($14.7\%$)[20]&19($7.1\%$)[12]\\
	High-L LMXBs	&49($30.2\%$)[19] &22($13.6\%$)[11]&10($6.2\%$)[5]\\
	Low-L LMXBs	&120($35.4\%$)[57]&63($18.6\%$)[37]&41($12.1\%$)[29]\\  
	\tableline	
	\end{tabular}
	\label{tab:numover}
\end{table*}	

\begin{figure*}[]
	\includegraphics[height=8cm,width=8cm,angle=0]{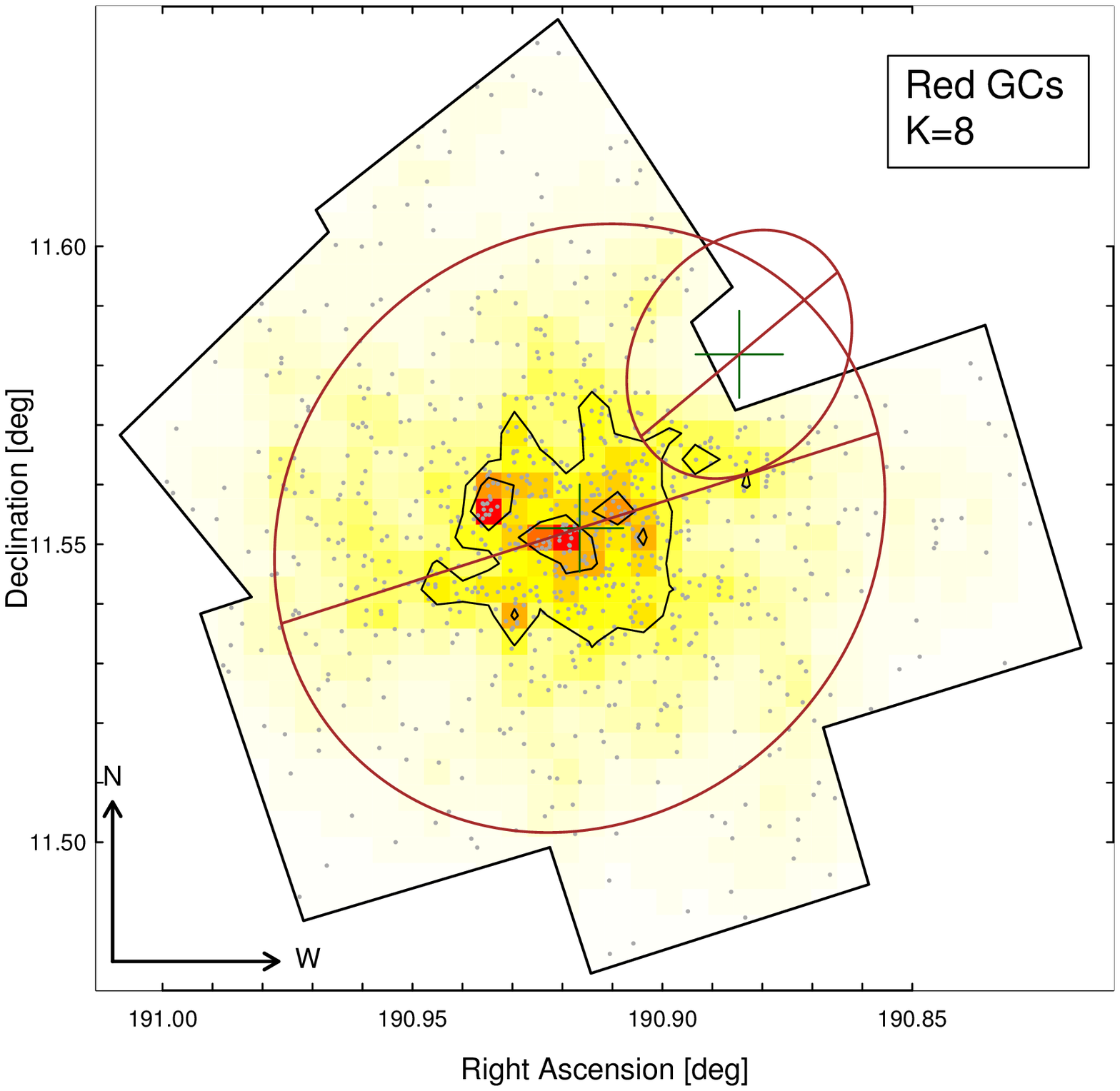}
	\includegraphics[height=8cm,width=8cm,angle=0]{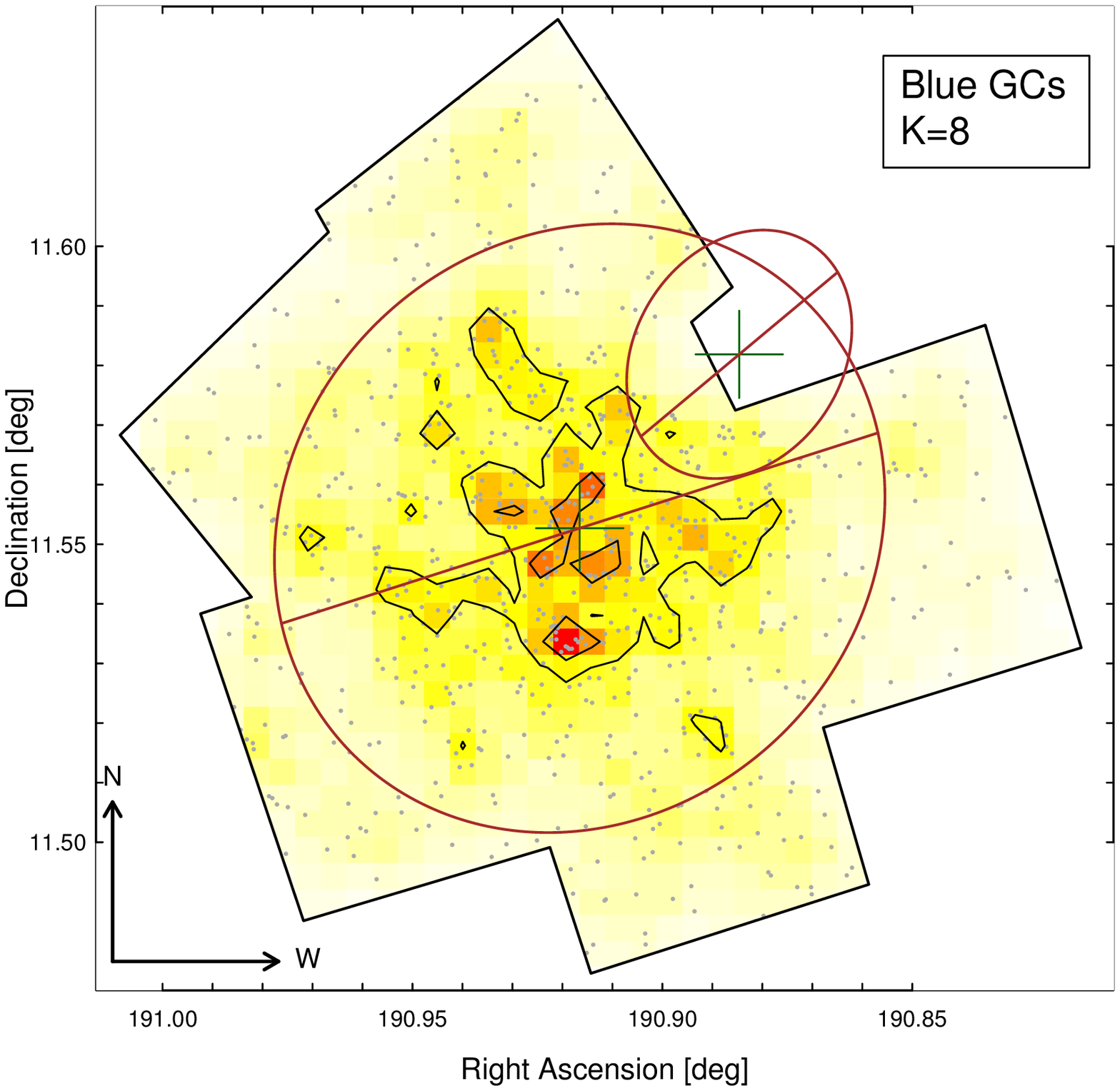}\\
	\includegraphics[height=8cm,width=8cm,angle=0]{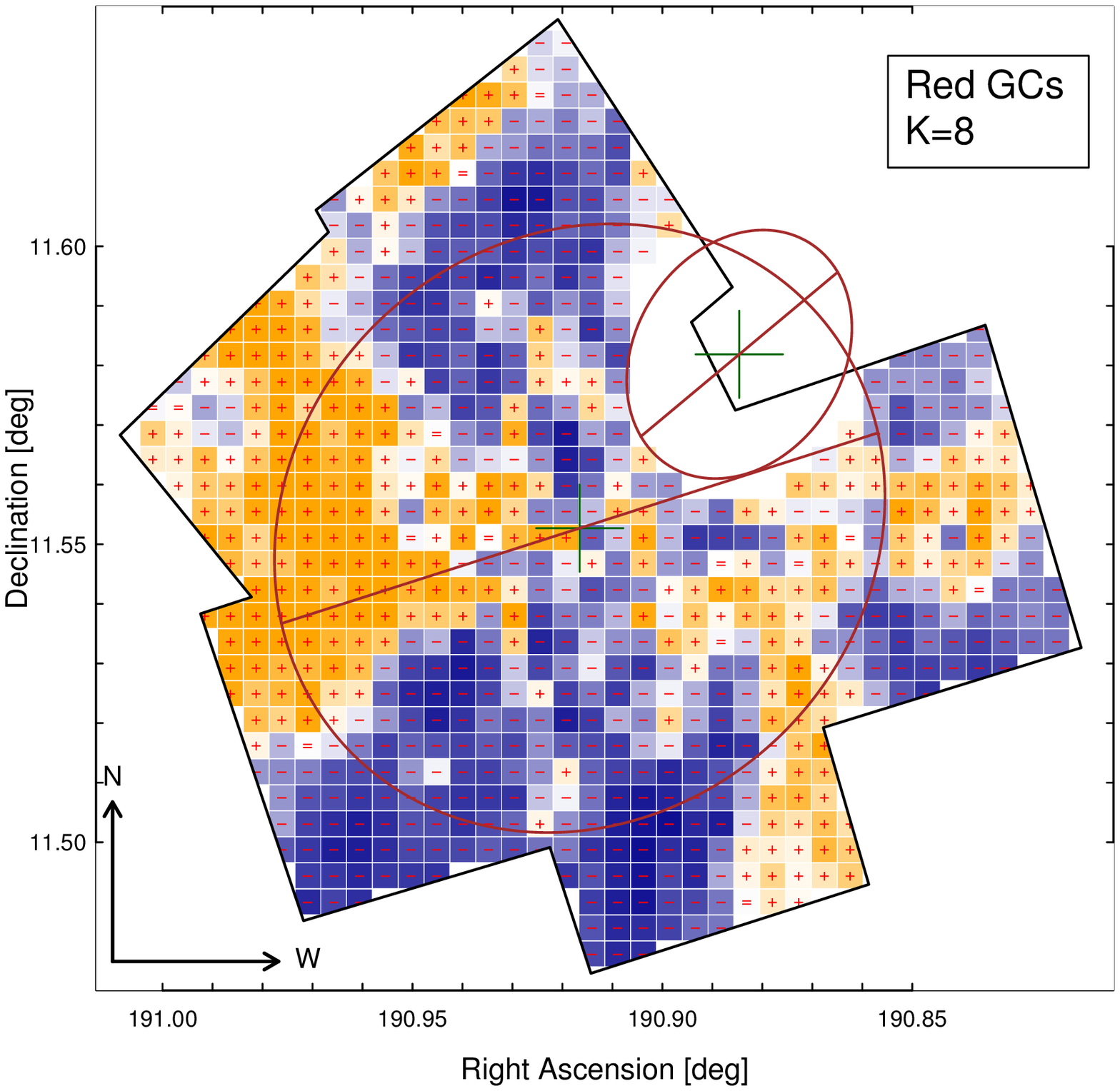}
	\includegraphics[height=8cm,width=8cm,angle=0]{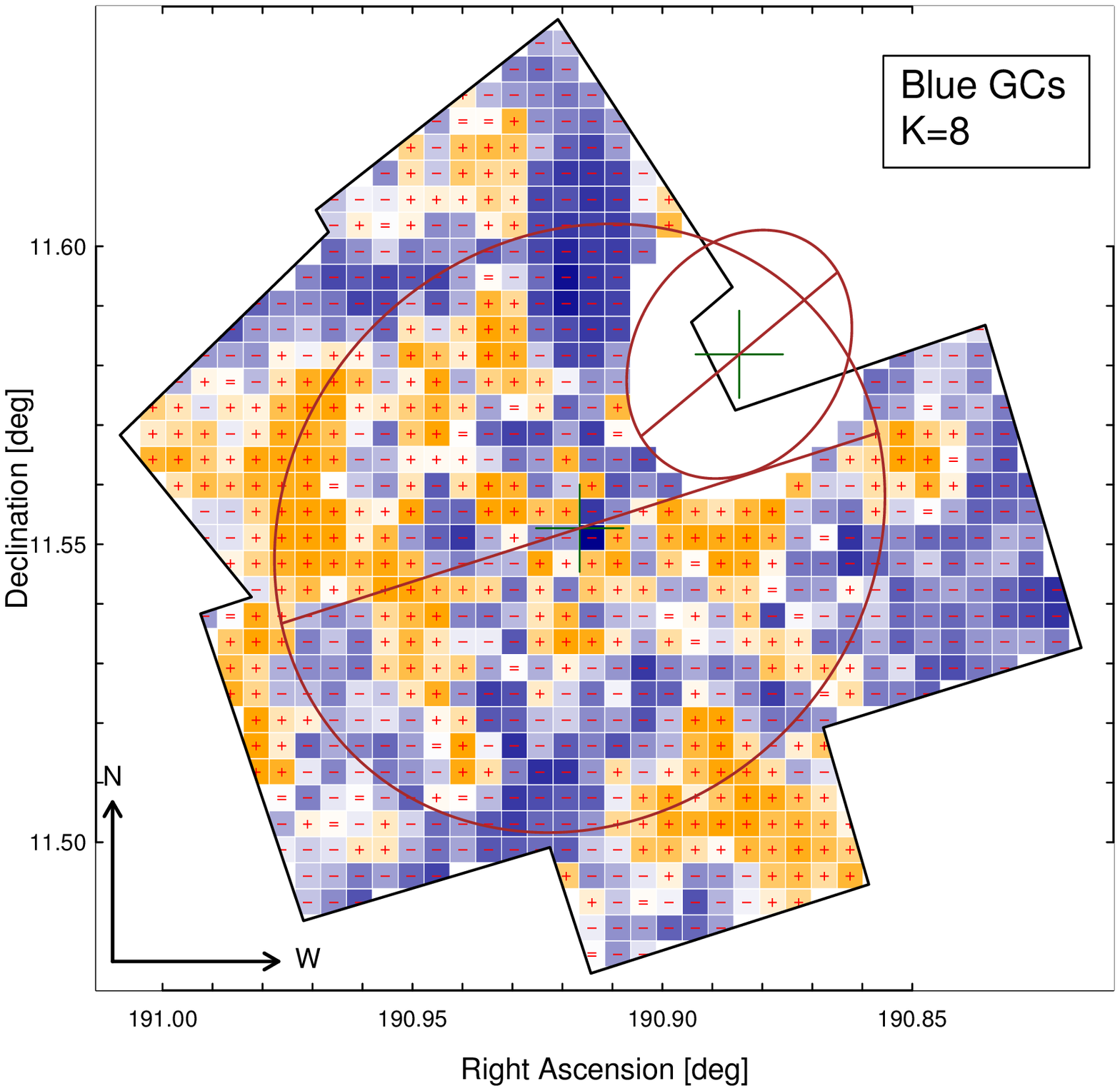}\\
	\caption{Upper panels: observed density maps of red (left) and blue (right) GCs samples for $K\!=\!8$. 
	Lower panels: residuals maps of red (left) and blue (right) samples obtained for $K\!=\!8$. The 
	small $+, -$ and $=$ signs within each pixel indicate positive, negative or null residuals respectively.
	In all plots the $D_{25}$ elliptical isophotes of both galaxies and the footprint of the HST 
	observations are shown for reference. The arbitrary isodensity contours in the upper plots are 
	only shown to highlight the position of the main over-densities.}
	\label{fig:redbluengc4649_gc}
\end{figure*}

The $K\!=\!8$ density maps of red and blue GC subsamples do not 
show significant differences in the location of the main over-densities (upper panels in 
Figure~\ref{fig:redbluengc4649_gc}), except for red GCs being more centrally concentrated 
(see also Figure~\ref{fig:2dmapsngc4649_gc}, center). The residuals map derived from the red 
GCs distribution (lower left panel in 
Figure~\ref{fig:redbluengc4649_gc}) highlights a large contiguous
region of positive residuals in the N-E quadrant, with two smaller regions of positive residuals located in 
the S-W quadrant. The $K\!=\!8$ 
residual map of blue GCs (lower right panel in Figure~\ref{fig:redbluengc4649_gc}) displays a 
more scattered distribution of positive residuals: a marginal enhancement corresponds to the 
region occupied by the large N-W positive residual structure seen in the red GCs map, but small significant 
clusters of positive residual pixels are located also in other regions. 

%This is also evident 
%in the plots of the residuals with significance $>\!1\sigma$ for both red and blue 
%GCs classes (lower panels in Figure~\ref{fig:redbluengc4649_gc}). 

\begin{figure*}[!h]
	\includegraphics[height=8cm,width=8cm,angle=0]{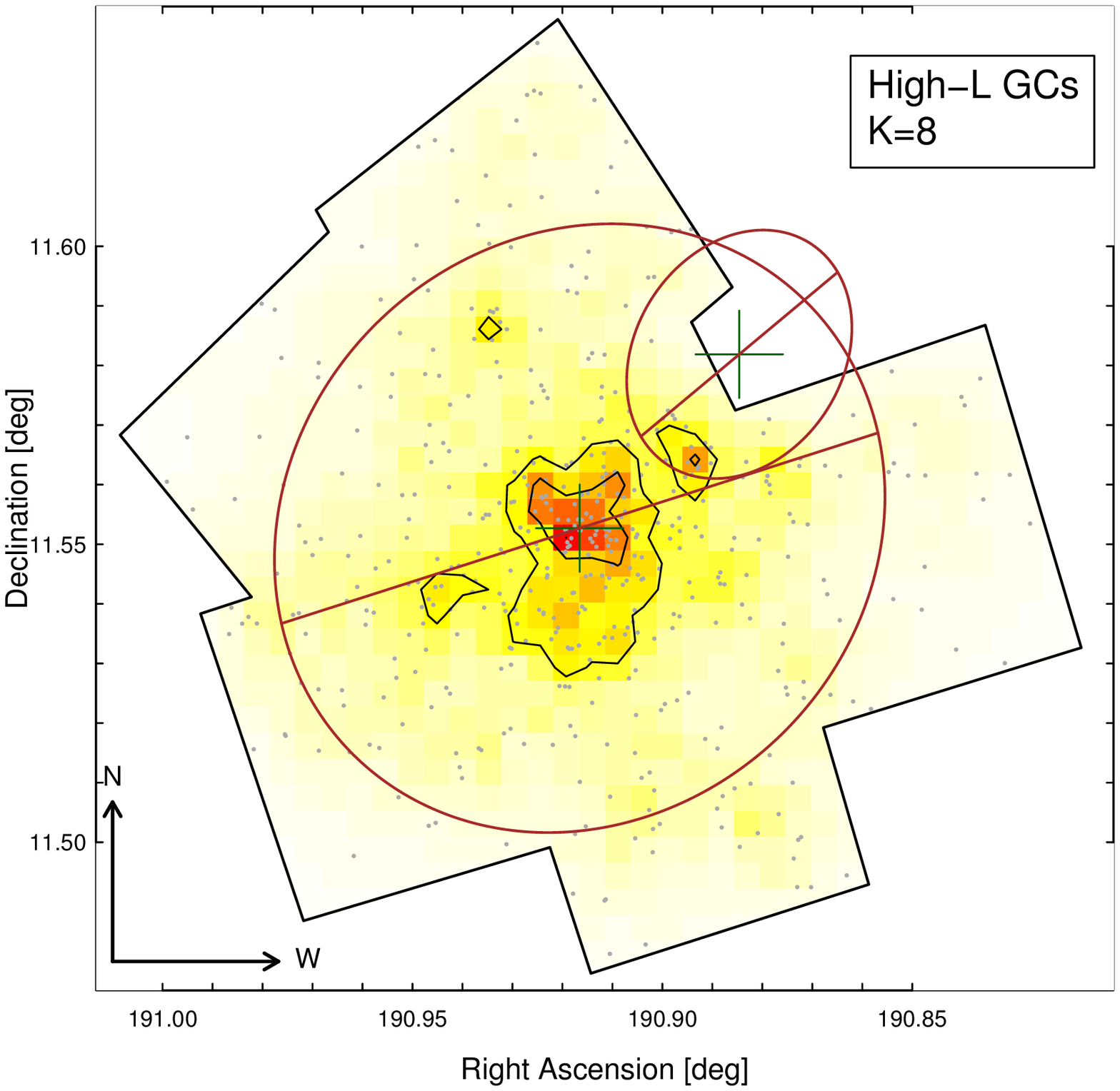}
	\includegraphics[height=8cm,width=8cm,angle=0]{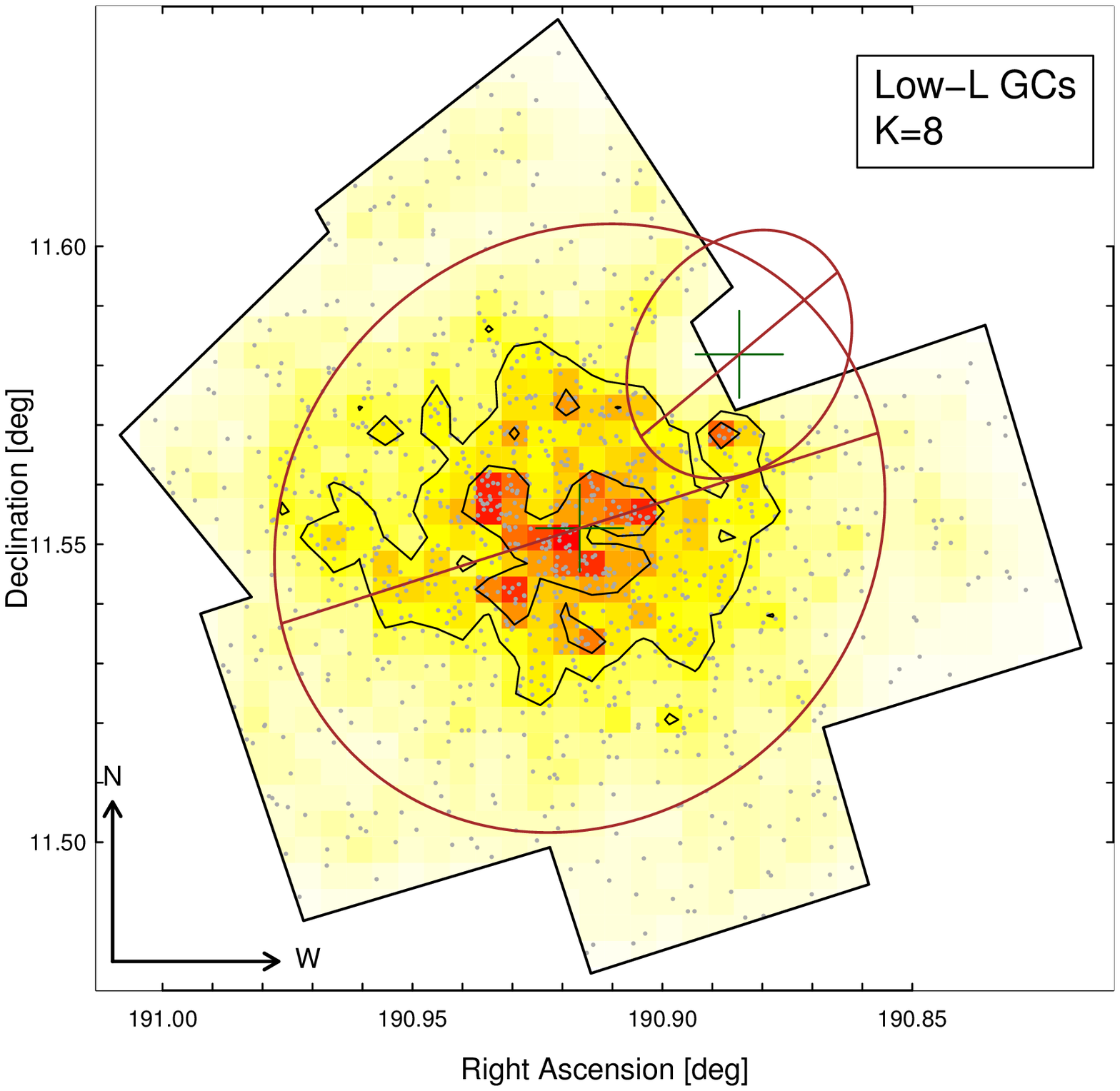}\\
	\includegraphics[height=8cm,width=8cm,angle=0]{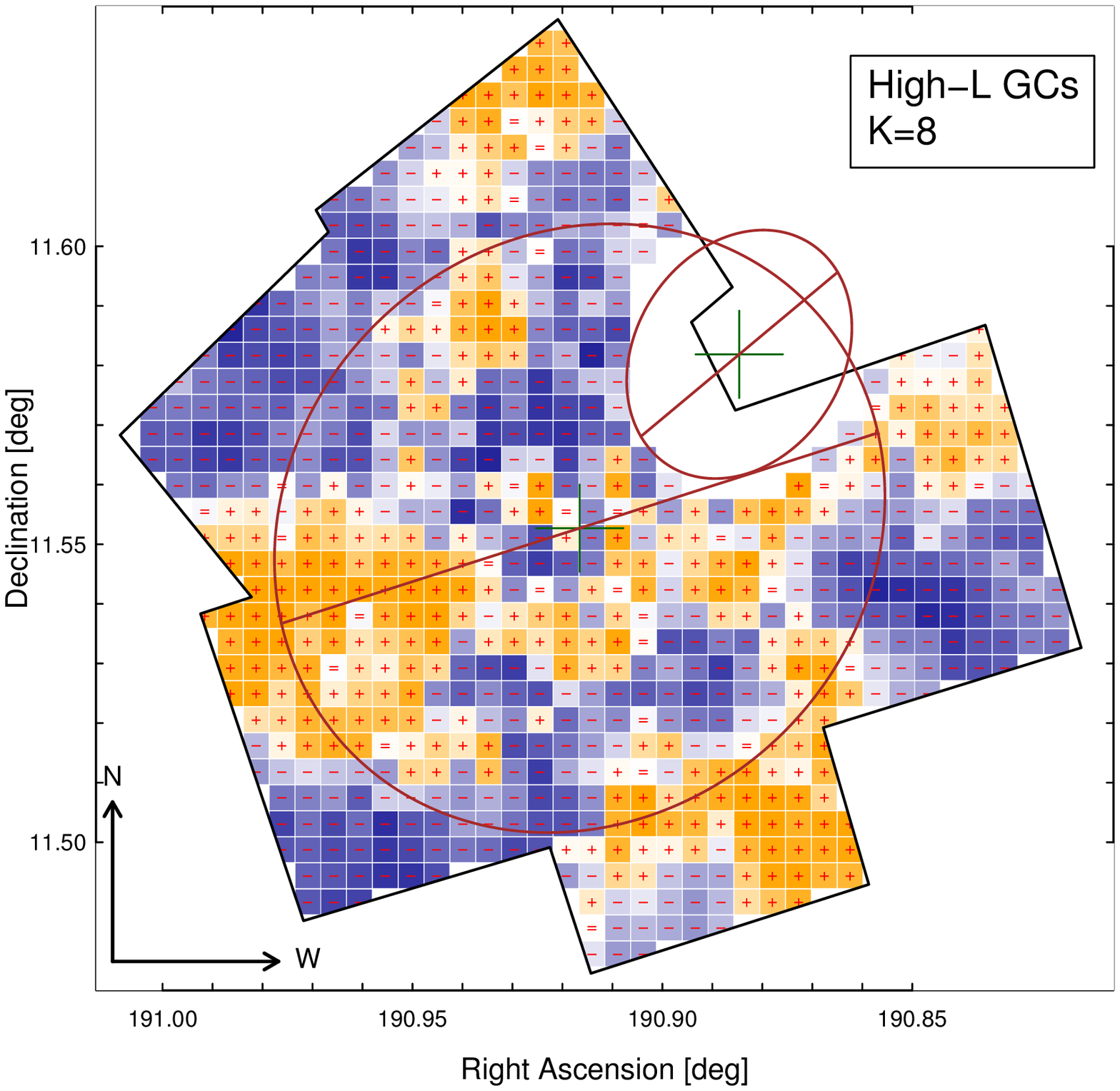}
	\includegraphics[height=8cm,width=8cm,angle=0]{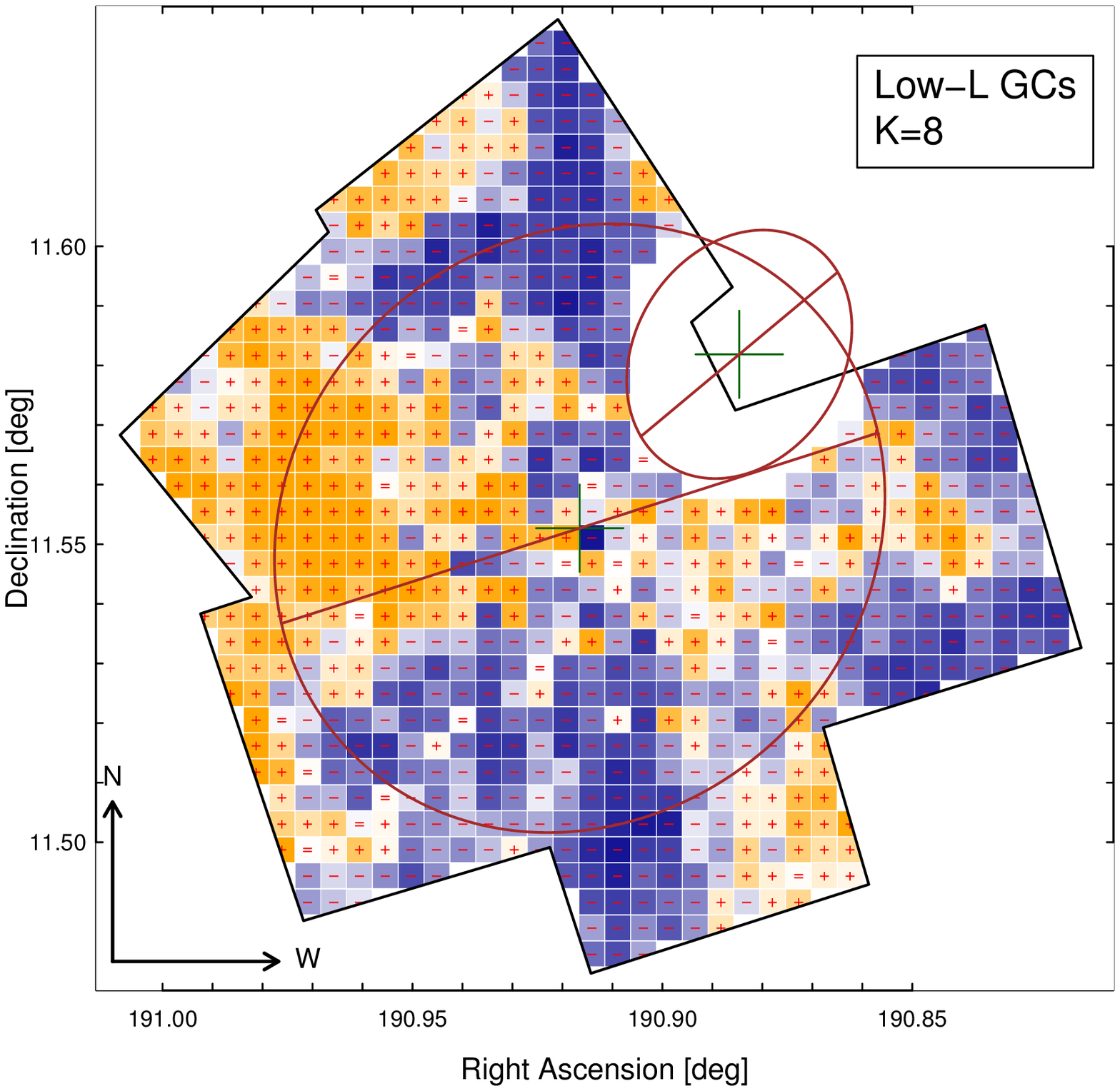}\\	
	\caption{Upper panels: observed density maps of the high-luminosity (left) GCs ($g\!\leq\!23$) and 
	low-luminosity (right) GCs ($g\!>\!23$) samples for $K\!=\!9$. Lower panels: 
	$K\!=\!8$ residual maps of high- (left) and low-luminosity (right) GCs samples. The 
	small $+, -$ and $=$ signs within each pixel indicate positive, negative or null residuals respectively.
	In all plots the $D_{25}$ elliptical isophotes of both galaxies and the footprint of the HST observations 
	are shown for reference. The arbitrary isodensity contours in the upper plots are 
	only shown to highlight the position of the main over-densities.}
	\label{fig:highlowlngc4649_gc}
\end{figure*}

Figure~\ref{fig:highlowlngc4649_gc} gives the corresponding results for the high and low 
luminosity GCs subsamples. The density distributions indicate that both luminosity 
classes peak in the center of the galaxy, with high-L GCs more centrally concentrated than low-L ones. 
The residual maps (lower panels in Figure~\ref{fig:highlowlngc4649_gc}) emphasize the 
differences between the two classes. The high-L GCs residual distribution is characterized by a 
significant cluster of spatially correlated pixels located along the E section of the major axis of the 
galaxy. Two other smaller structures are visible in the S-W quadrant within and outside of the 
$D_{25}$ isophote respectively. The residual map of the low-L GCs differs significantly from the 
high-L map in the E side: a large group of nearby pixels with highly significant positive residuals
is located almost entirely in the N-E quadrant. This structures occupies a region devoid of positive
high-L pixels and overlaps only partially with the high-L main over-density structure aligned along the 
major galaxy axis in the E direction. 

%The differences between the distribution of positive residual pixels 
%for the two GCs luminosity classes are clearly visible in the lower panels in 
%Figure~\ref{fig:highlowlngc4649_gc}, where only pixels associated to residuals with 
%significance $>\!1\sigma$ are shown.

Following~\cite{dabrusco2013}, we have verified that the differences 
between the spatial distributions of Red and Blue GCs, and high-L and low-L GCs are not the results 
of a particular choice of the thresholds values by reconstructing the density and residual
maps of the two classes for several values of the thresholds, and checking that the qualitative results
do not change. For the color classes, we have used ten regularly spaced threshold values around
the threshold value $g\!-\!z\!=\!1.18$ in the interval $g\!-\!z\!=\![1, 1.3]$. The results are consistent 
throughout this range of colors. For values outside this interval, the significance of the residual 
map for one of the two classes degrades rapidly because of the small number of GCs in that class.
The same qualitative conclusions discussed above for the characterization of the spatial distributions 
of low and high luminosity classes of GCs in NGC4649 hold true using ten regularly spaced values of the 
magnitude threshold within the interval $g\!=\![22.5, 23.5]$. Other values of the $g$ magnitude threshold 
have not been used since they would generate luminosity classes too unbalanced to 
correctly estimate the significance of the results.

\subsection{Density and residual maps of LMXBs}
\label{subsec:lmxbdensity}

\begin{figure*}[]
	\includegraphics[height=6cm,width=6cm,angle=0]{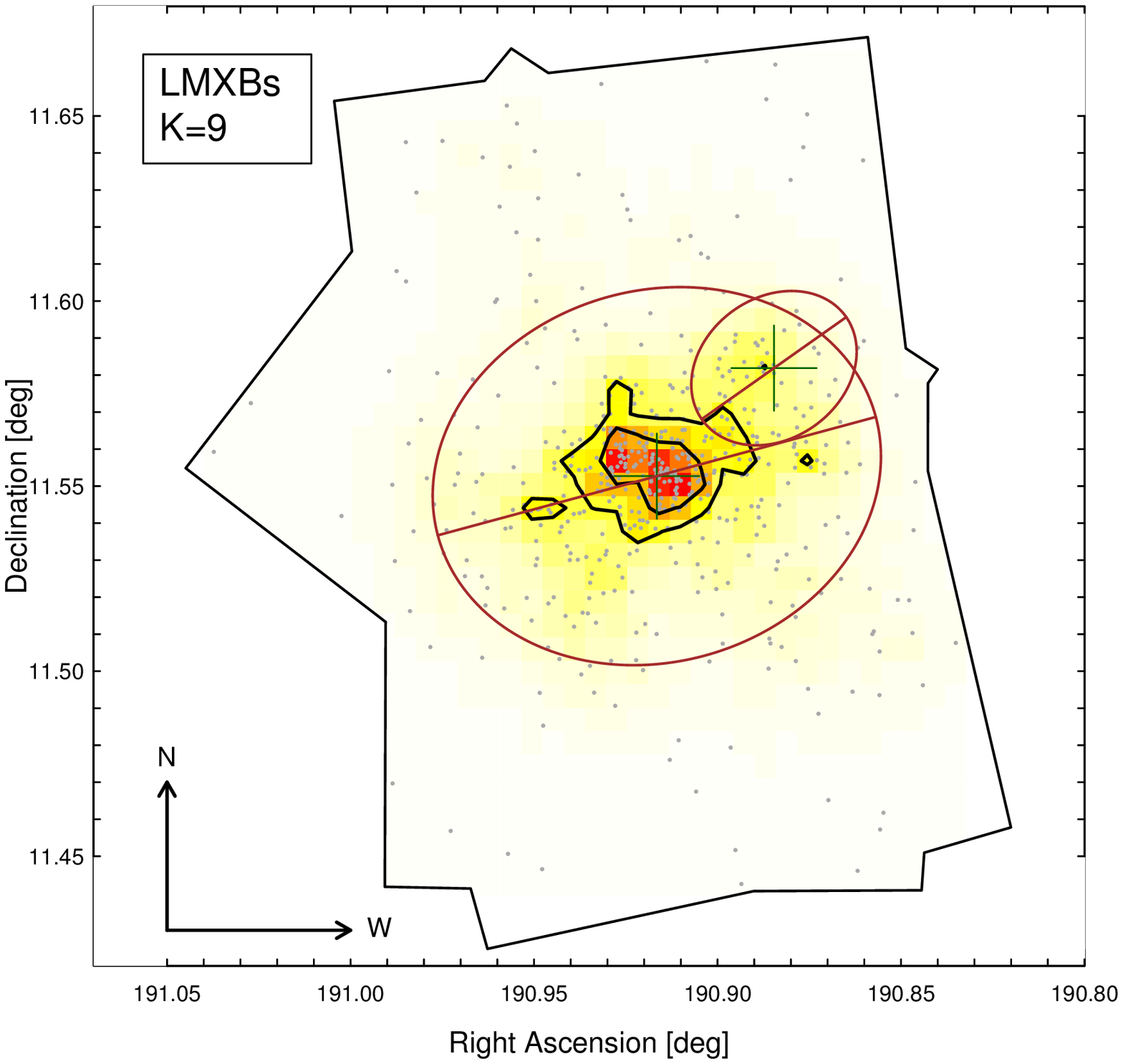}
	\includegraphics[height=6cm,width=6cm,angle=0]{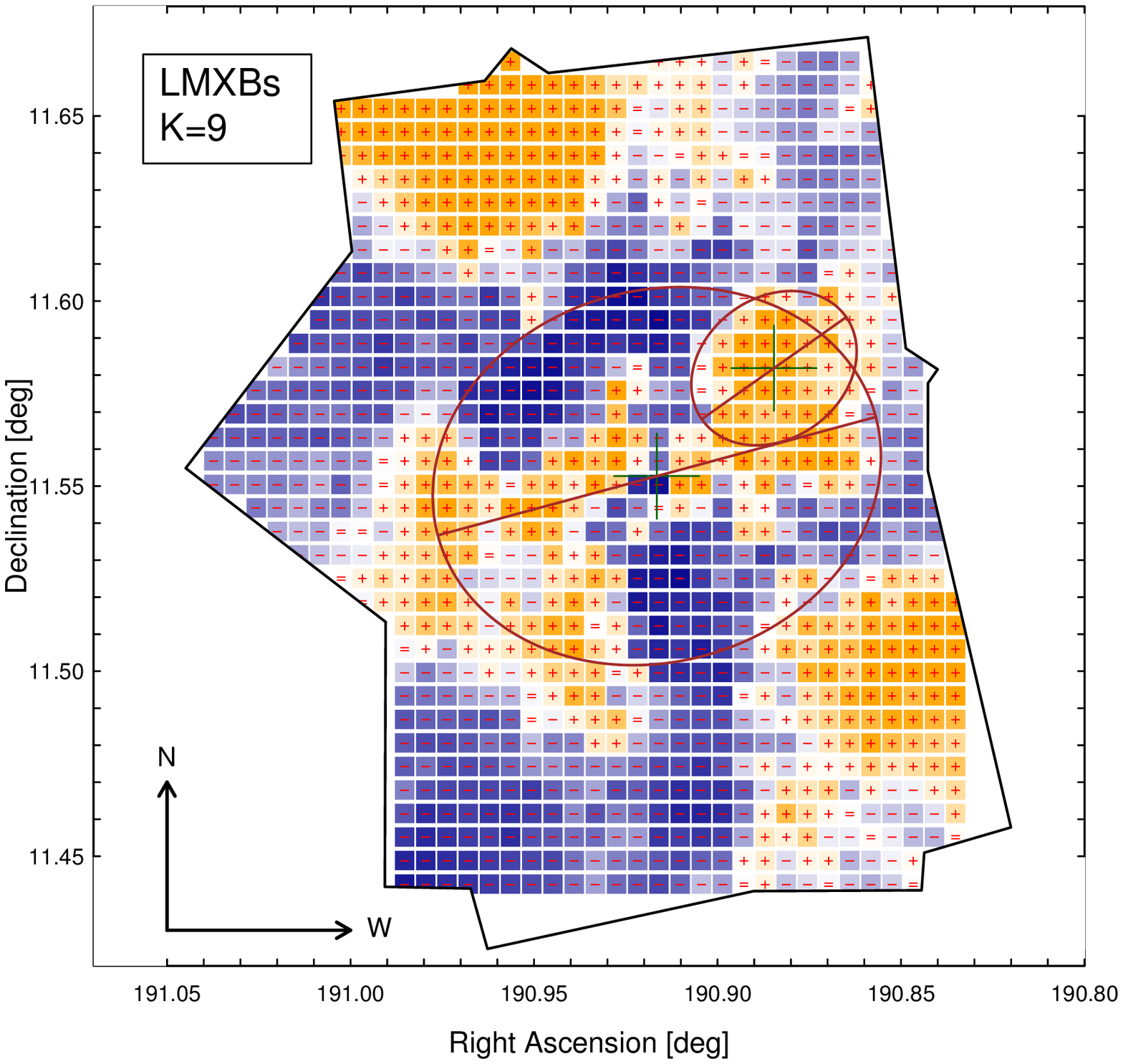}
	\includegraphics[height=6cm,width=6cm,angle=0]{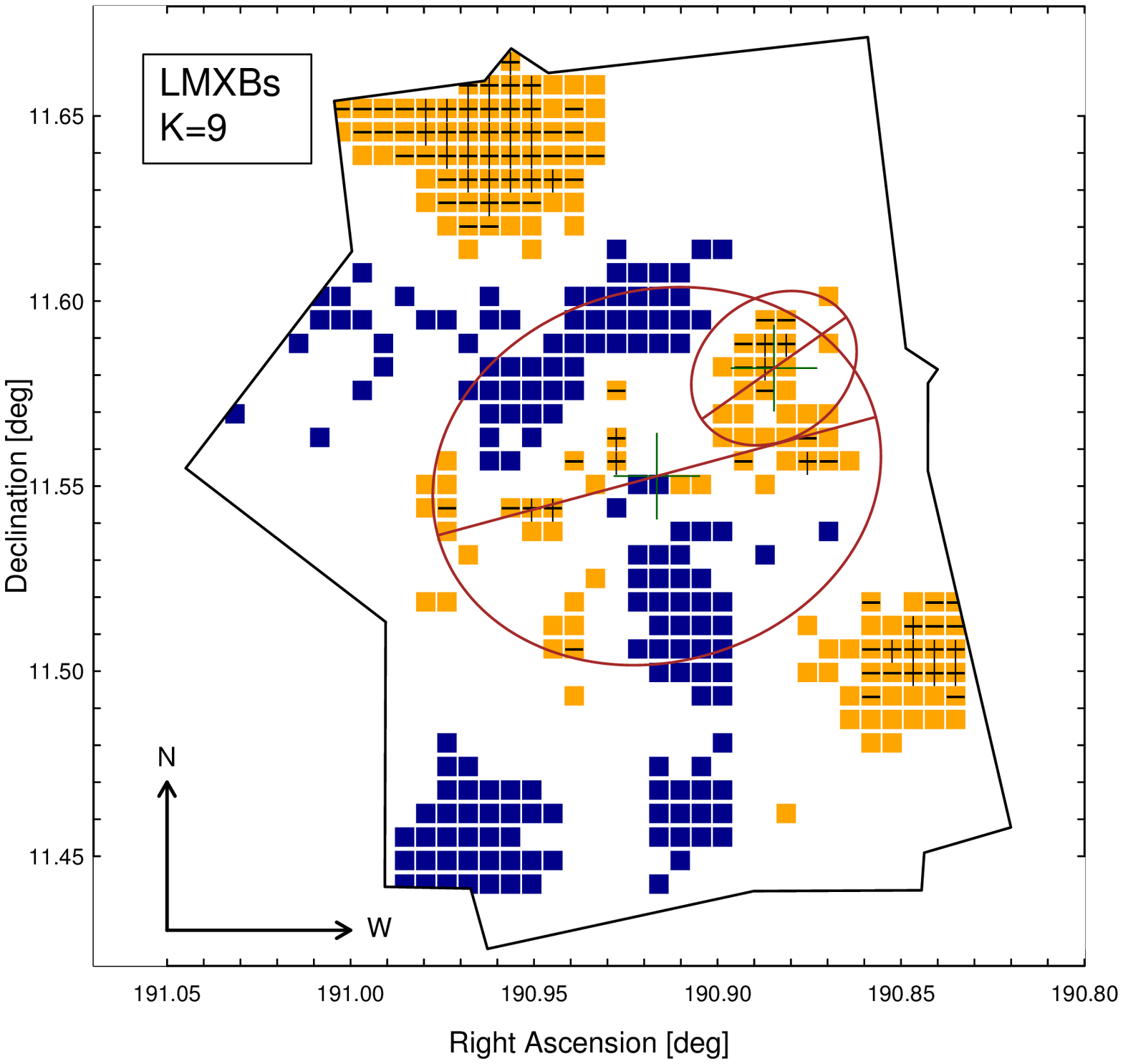}	
	\caption{Left: $K\!=\!9$ density maps of the sample of LMXBs in NGC4649. Arbitrary isodensity contours 
	show the higher-density regions in each map. The arbitrary isodensity contours are 
	only shown to highlight the position of the main over-densities.
	Center: $K\!=\!9$ residual map of the entire sample of 
	LMXBs detected in the {\it Chandra} fields centered on NGC4649. Pixels are color-coded 
	according to the number of $\sigma$ the pixel deviates from the average. Darker colors indicate larger
	residuals: blue, negative; orange, positive. The small $+, -$ and $=$ signs within each 
	pixel indicate positive, negative or null residuals respectively. 
	Right: Positions of the $K\!=\!9$ residuals with significance larger than 
	1$\sigma$, 2$\sigma$ and 3$\sigma$ obtained from the residual maps derived from the 
	distribution of the whole catalog of LMXBs in NGC4649. All the negative residuals (blue pixels)
	have significance between 1 and 2 $\sigma$. Positive (orange) pixels $\!>\!2\sigma$ are indicated 
	with a horizontal line; $\!>\!3\sigma$ with a cross. The footprint of the {\it Chandra} observations
	used to extract the catalog of LMXBs and the $D_{25}$ isophotes of both NGC4649 and NGC4647 are 
	shown in all plots.}
	\label{fig:2dmapsngc4649_lmxb}
\end{figure*}

The distribution of LMXBs is centrally concentrated (Figure~\ref{fig:2dmapsngc4649_lmxb}, left), except 
for an over-density located within the $D_{25}$ 
elliptical isophote of the companion spiral galaxy NGC4647 and a less significant enhancement in the S-E 
quadrant within the $D_{25}$ NGC4649 isophote. The residual map enhances these over-densities 
(Figure~\ref{fig:2dmapsngc4649_lmxb}, center, right).
A significant association of positive residuals pixels can be seen along the major axis of NGC4649. 
Two other significant regions composed of multiple spatially clustered positive residual pixels are 
visible in the N-E and S-W corners of the footprint of the {\it Chandra} observations. Although 
the footprint of the~\cite{strader2012} does not include the regions of the 
{\it Chandra} field occupied by these two structures, three X-ray sources in the N-E over-density and 
four in the S-W density enhancement have been associated by~\cite{luo2013} 
to GCs from the catalog of~\cite{lee2008}. The remaining X-ray sources in the two 
structures do not have an optical counterpart that could be identified by our cross-correlation with 
the Nasa Extragalactic Database\footnote{http://ned.ipac.caltech.edu/} (NED), Simbad Astronomical
Database\footnote{http://simbad.u-strasbg.fr/simbad/}, Sloan Digital Sky 
Survey\footnote{http://www.sdss3.org/dr10/} 
(SDSS) DR10, Two Micron All Sky Survey\footnote{http://www.ipac.caltech.edu/2mass/} (2MASS) and 
Wide-Field Infrared Survey Explorer\footnote{http://wise2.ipac.caltech.edu/docs/release/allwise/} (WISE) 
ALLWise archives. Nine Ultra-Luminous X-ray 
sources (ULXs) with $L_{X}\!>\!10^{39}$ ergs s$^{-1}$ were identified by~\cite{luo2013}, 
but they are not located within the boundaries of the N-E and S-W structures.
The presence of LMXBs associated to GCs~\citep{lee2008} suggests that these two over-densities 
could be the remnants of past interactions
between NGC4649 and small accreted satellite galaxies. However, considering the background AGN density 
calculated in~\cite{luo2013}, we cannot rule out contribution from background 
AGNs.

Differences are found in the distribution of residuals for field vs GCs-LMXBs and low and high X-ray
luminosities LMXBs. These will be discussed in Section~\ref{sec:discussion}.

\subsection{Comparison with the distribution of diffuse stellar light}
\label{subset:diffuselight}

We have searched for structures in the distribution of the diffuse stellar light of NGC4649
observed in the HST data used by~\cite{strader2012} that could be
spatially overlapping with the over-density structures observed in the spatial distribution of GCs. 
We have fitted and subtracted elliptical isophotal models to the images of NGC4649 galaxy in both 
$g$ and $z$ filters, without finding significant residuals that may be indication of merging
(cp. with the results discussed by~\cite{tal2009,janowiecki2010}). 
Also, we searched for locally enhanced stellar formation in the color map of NGC4649 
obtained by combining the HST g and z, using the method based on the distribution of 
the pixels images in the color-magnitude (CM) diagram~\citep[see][]{degrijs2006}. 
We do not observe significant differences in the distribution of pixels in the CM diagram 
derived for different azimuthal regions of the galaxy containing the GC over-density structures.
The only feature in the pixel CM diagram is the ``blue'' cloud of pixels in the region of NGC4649 
overlapping with NGC4647, already observed by~\citep{degrijs2006}).
While our results do not suggest the existence of significant deviations from a smooth model of 
the diffuse light and color distributions in NGC4649, more detailed analysis will be necessary to 
confirm the lack of faint structures in the diffuse light 2D model of this galaxy.
%Also, we do not detect significant color differences~\citep[e.g.][]{degrijs2006} correlated with the 
%GC spatial anisotropies.

\begin{figure}[]
	\includegraphics[height=8cm,width=8cm,angle=0]{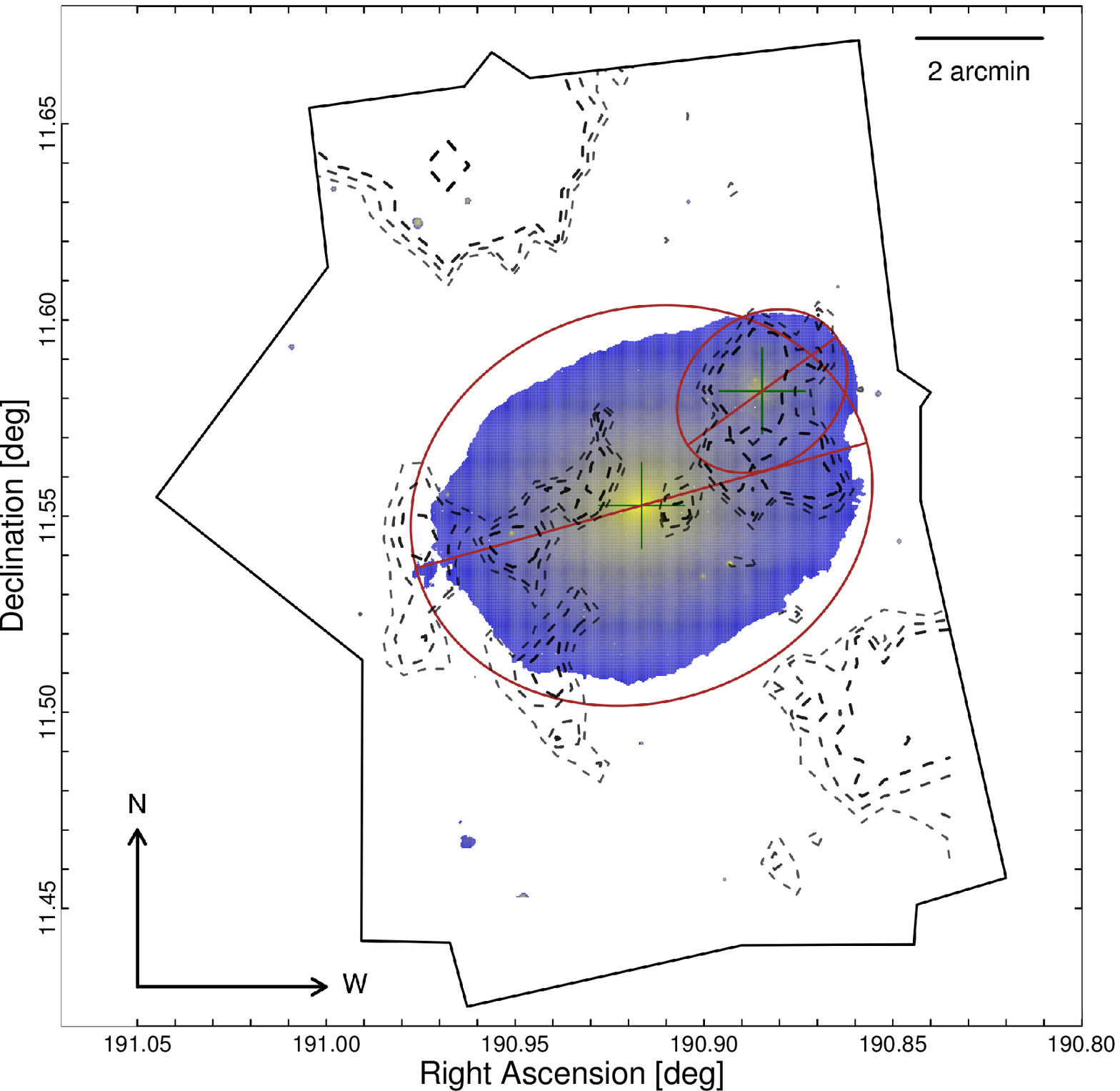}		
	\caption{Stellar mass density map for NGC4649 and NGC4647, as calculated by~\cite{mineo2013}.
	Isodensity contours from the residual map distribution of all LMXBs from 
	Figure~\ref{fig:2dmapsngc4649_lmxb} are also shown.}
        \label{fig:stellarmassdensity_ngc4649}
\end{figure}

In Figure~\ref{fig:stellarmassdensity_ngc4649} 
we plot the isodensity contours
of the over-density structures observed in the spatial distribution of LMXBs in NGC4649 over the stellar
mass map from~\citep{mineo2013}. This stellar mass map
is smooth, with a steep positive gradient towards the core of NGC4649. The
only significant enhancement is associated to the central region of the companion galaxy NGC4647, 
where some granularity, possibly due to the spiral arms, is visible (see Figure~10 
in~\citep{mineo2013}). The position of the positive 
residual structures on the E side of NGC4649 does not correlate with any feature in the stellar 
mass map, while on the W side the LMXBs over-density contours clearly follows the stellar mass
density enhancement associated to the spiral galaxy NGC4647. The different spatial 
resolutions of the residual maps of the distribution of LMXBs in NGC4649 and of the stellar mass density map
do not permit a direct comparison of the shape of the LMXBs over-density contours to the 
position of the granularities of the map though.

\section{Discussion}
\label{sec:discussion}

Our analysis of the two-dimensional projected distribution of the GC and LMXB populations of the giant 
Virgo elliptical NGC4649 (M60) confirms a well-known feature of this 
GC system, that red GCs are more centrally concentrated than blue GCs, as usually observed in 
elliptical galaxies (e.g.,~\citealt{brodie2006}; see also~\citealt{mineo2013} for NGC4649), but also 
finds unexpected significant and complex 2D asymmetries in their projected distributions.

\subsection{The GC population of NGC4649}
\label{subsec:gcngc4649}

NGC4649 is the third most luminous galaxy in the Virgo Cluster, and resides in a galaxy-dense 
environment, where gravitational interactions and accretion of satellites may be frequent.
Kinematics evidence, suggesting a merging and accretion past for this galaxy, was recently
found in studies of Planetary Nebulae (PN) and GCs~\citep{teodorescu2011,das2011,coccato2013}.
In particular,~\cite{das2011} show that at galactocentric radii larger than 12 kpc, PNs and GCs
may belong to separate dynamical systems. A similar conclusion was reached by~\cite{coccato2013},
who propose either tidal stripping of GCs for less massive companion galaxies, or a combination of 
multiple mergers and dwarf galaxy accretion events to explain the observed kinematics. Besides 
experiencing continuing evolution at the outer radii via interaction with and accretion of companions, NGC4649
itself may be the result of a major dry merger. This is suggested by recent kinematical measurements
of the diffuse stellar light that revealed disk-like outer rotation, as it may have stemmed from a  
massive lenticular galaxy progenitor~\citep{arnold2013}.

\begin{figure*}[]
	\includegraphics[height=8cm,width=8cm,angle=0]{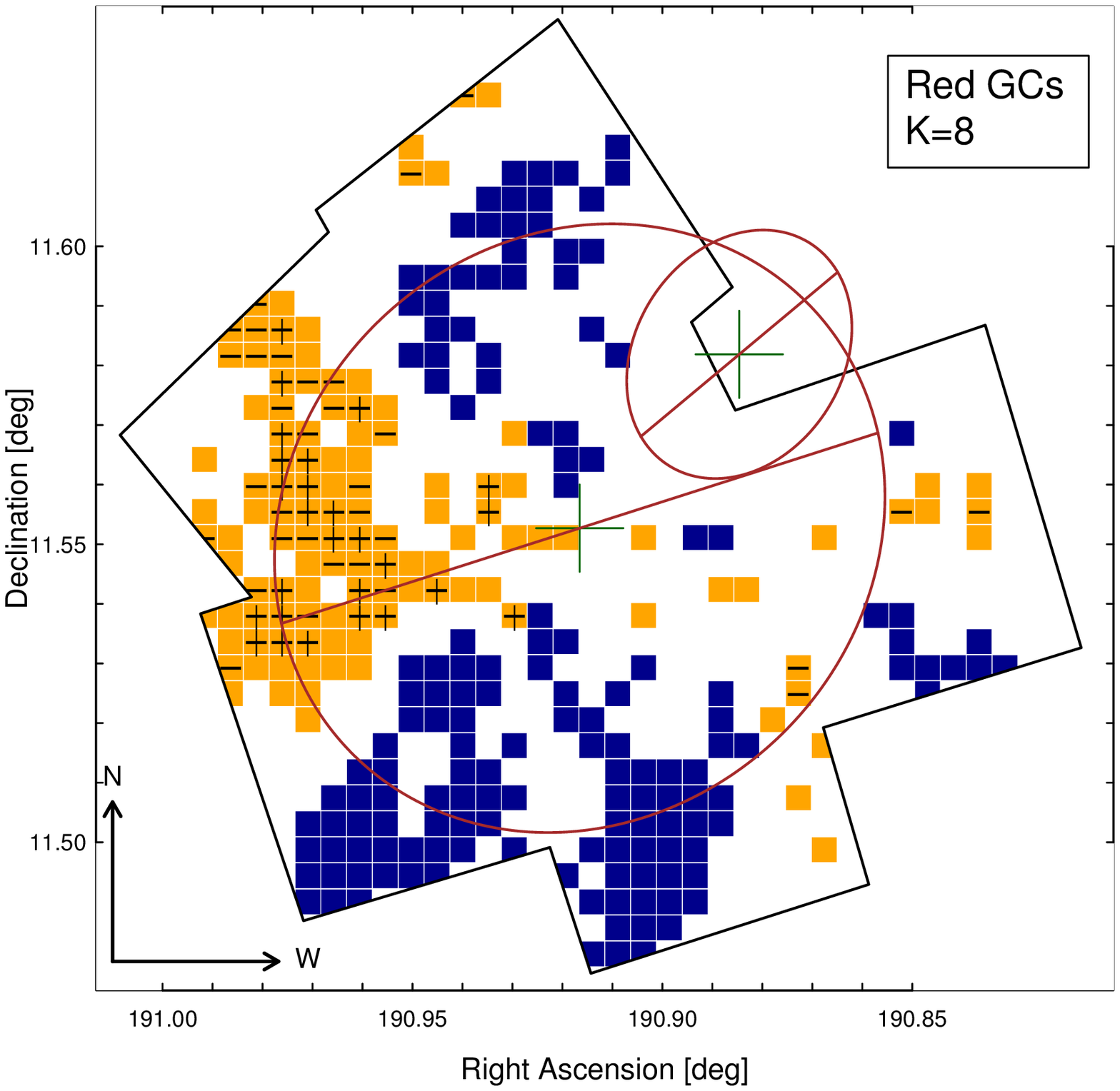}	
	\includegraphics[height=8cm,width=8cm,angle=0]{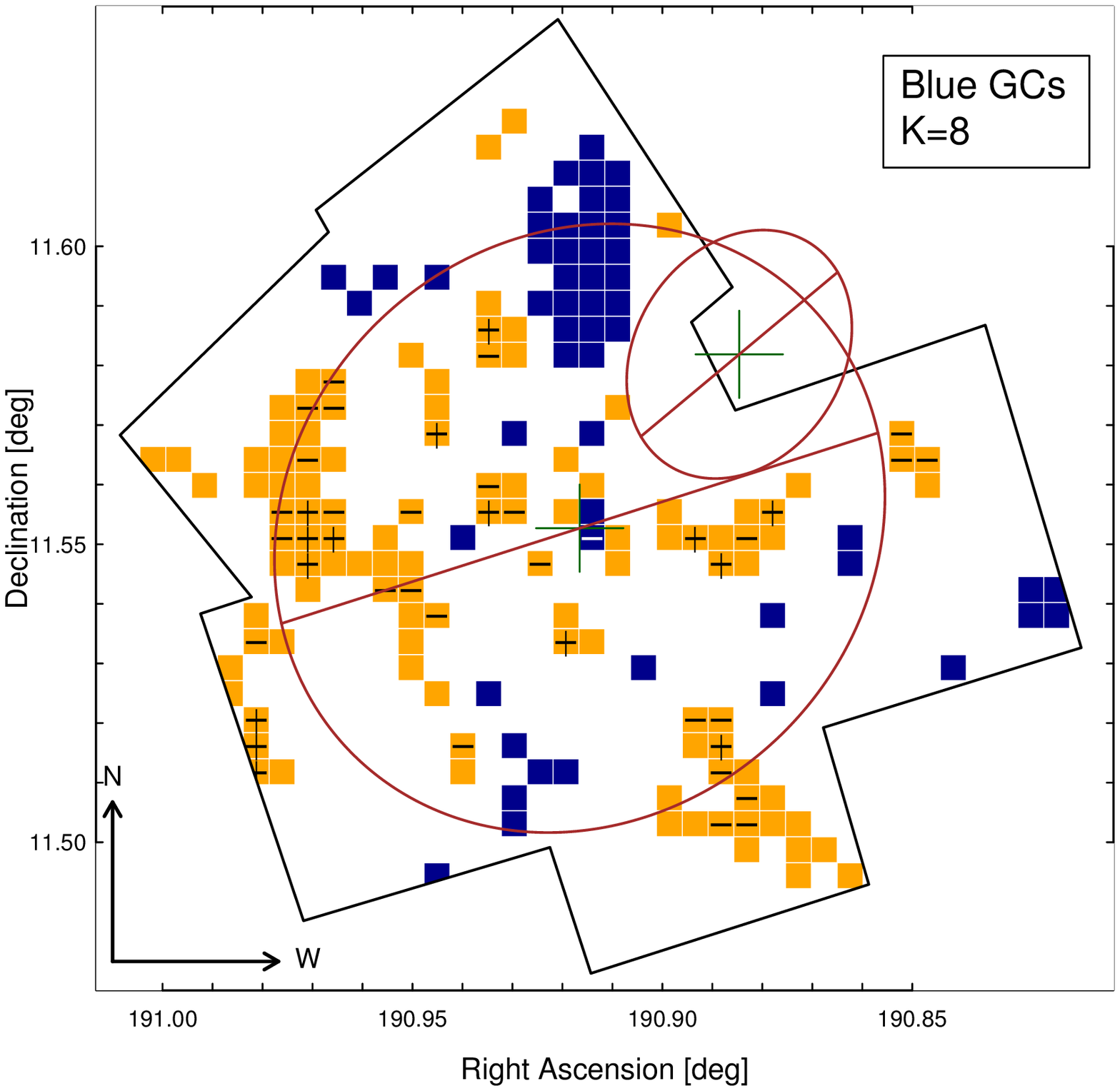}\\	
	\includegraphics[height=8cm,width=8cm,angle=0]{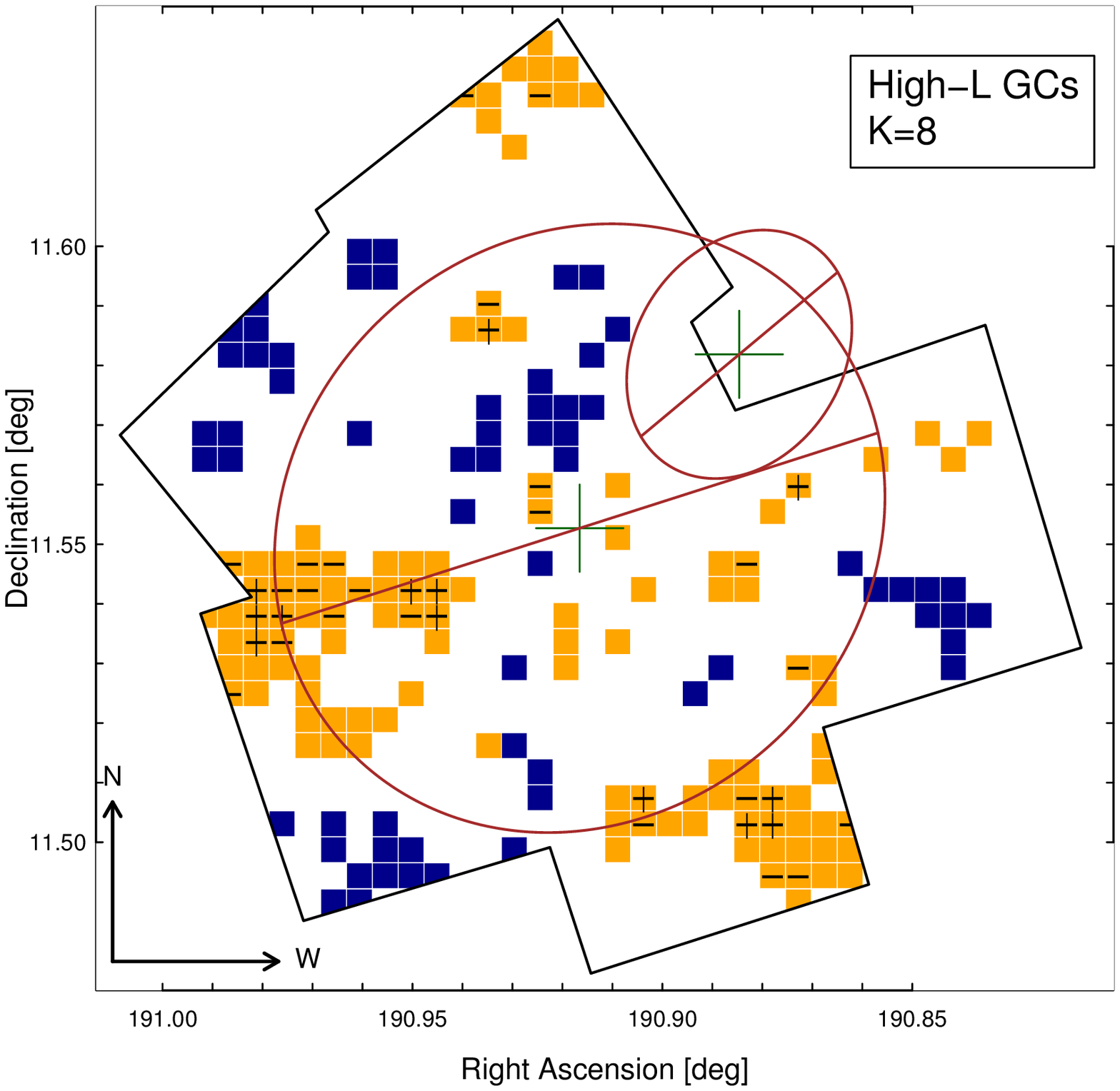}	
	\includegraphics[height=8cm,width=8cm,angle=0]{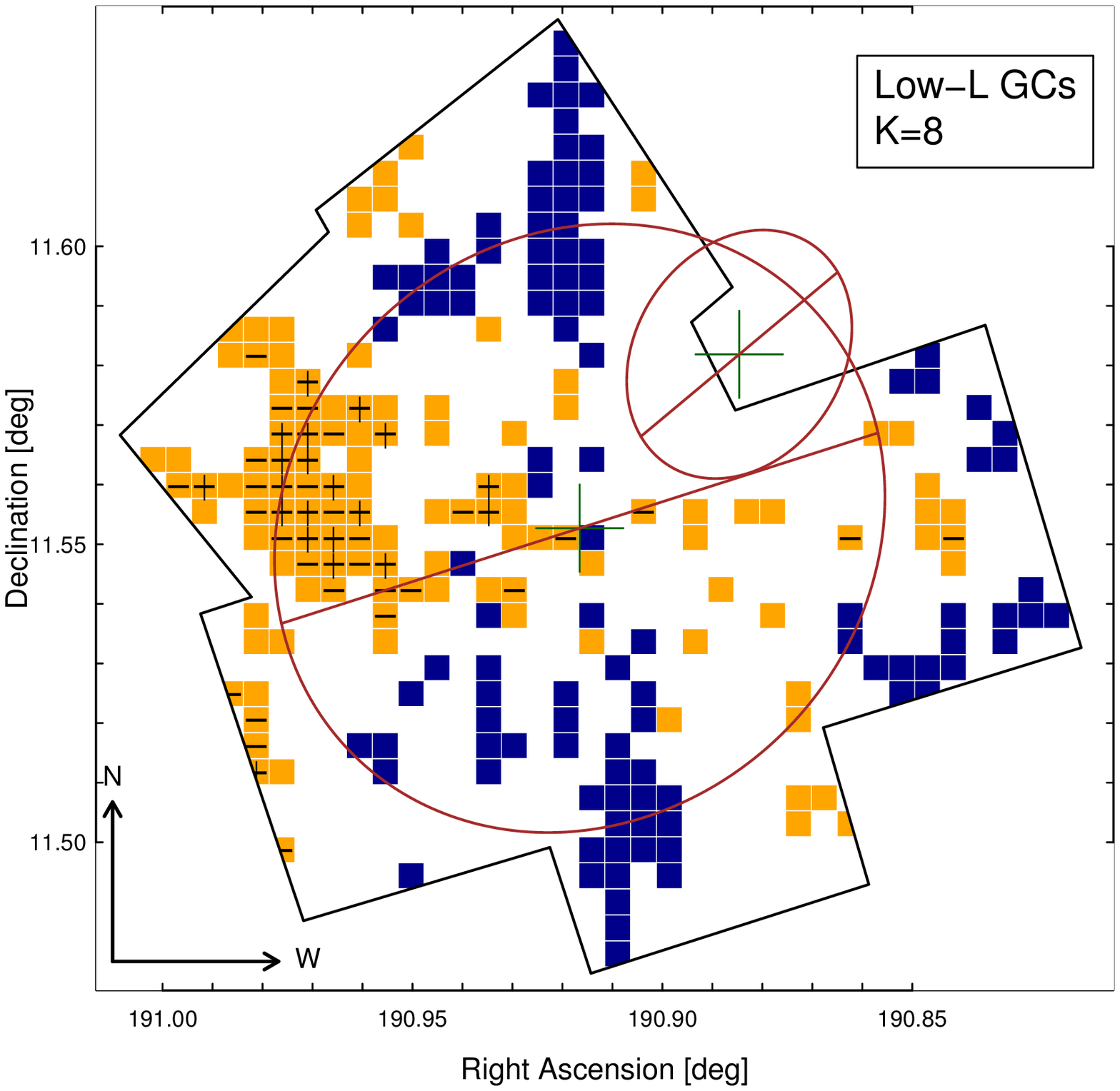}\\		
	\caption{Upper panels: positions of the $\!>\!1\sigma$ residuals for blue (left) and red (right) samples of 
	GCs obtained for $K\!=\!8$. Positive (orange) pixels $\!>\!2\sigma$ are indicated 
	with a horizontal line. Lower panels: positions of the $\!>\!1\sigma$ residuals for high- (left) and low-luminosity 
	(right) samples of GCs obtained for $K\!=\!8$. Positive (orange) pixels $\!>\!2\sigma$ are indicated 
	with a horizontal line; $\!>\!3\sigma$ with a cross. In all plots the $D_{25}$ elliptical isophotes 
	of both galaxies and the footprint of the HST observations are shown for reference.}
	\label{fig:sigma_pixels_maps_gc}
\end{figure*}

The anisotropies we find in the 2D distribution of GCs may be a different pointer to
this complex evolution. We have found that in NGC4649, 
the 2D distribution of GCs shows strong positive residuals along the eastern major axis of the galaxy, 
with a northward arc-like curvature beginning at a galacto-centric radius of $\sim\!4$ kpc and extending 
out to $\sim\!15$ kpc. 
This large-scale feature is more prominent in the distribution of red GCs, but can still be seen in the 
2D distribution of blue GCs, which is overall noisier. High-luminosity GCs tend to concentrate at the 
southern end of this feature, while low-luminosity GCs are found in the northern portion. These trends 
and differences are highlighted in Figure~\ref{fig:sigma_pixels_maps_gc}, which displays the most 
significant single pixels ($>\!1\sigma,2\sigma,3\sigma$) of each residual distribution. The extreme 
pixel distributions of red GCs (upper left) and low-luminosity GCs (lower right) are similar, and differ from 
those of high-L GCs. The
reason for this similarity resides in the choice of the particular magnitude threshold used to 
define luminosity classes (see discussion in Section~\ref{sec:data} and Table~\ref{tab:summary}). 
With the $g\!\leq\!23$ threshold, red and blue GCs comprise $\sim\!76\%$ and $\sim\!65\%$ of 
low-L GCs, respectively. Using $g\!\leq\!22$ as a threshold instead, the residual
maps of low-L and high-L GCs are more similar, although high-L and low-L still are spatially
segregated in the N side of the ``arc''. However, this magnitude threshold yields 
140 high-L GCs and 1463 low-L GCs.

The color and luminosity differences suggest that we may not simply be witnessing the disruption of a satellite dwarf 
galaxy, and observing the stream of its GC system. Perhaps some additional GC formation resulted from this 
merger, increasing the population of red, high-metallicity GCs. GC formation during galaxy merging has been 
suggested as an important mechanism for the assembly of GC systems, especially in the case of large 
elliptical galaxies~\citep{ashman1992}, and is supported by high-resolution simulations of galaxy 
mergers~\citep{bournaud2008}. Numerous young massive stellar clusters, which may evolve 
into GCs, are for example seen in the advanced merger remnant NGC7252~\citep{bastian2013}. This 
picture is being confirmed by other observational studies. A kinematical study of the GC system 
of the central galaxy of the Virgo Cluster, M87, 
shows the presence of significant sub-structures, pointing to active galaxy assembly~\citep{strader2011}. 
In the Virgo elliptical NGC4365, three separate rotating GC systems have been reported associated with the 
three color-families of GCs, in addition to a stream-like system~\citep{blom2012}.

With the exception of the large-scale arc structure observed in NGC4261~\citep{dabrusco2013}, little 
evidence is available in the literature about the 2D spatial distribution of GCs in early-type
galaxies, let alone about the differences in the spatial distributions of GC color and luminosity 
classes due to major mergers.~\cite{romanowsky2012} discussed the existence of a phase space shell 
composed of GCs located in the inner halo of M87, which is possibly due to a significant 
merger, but there is no evidence of similar structures in the GC spatial distribution in M87.
Based on these results, we deduce that the differences in the spatial distribution of high- and 
low-luminosity GCs in the N section of the NGC4649 arc suggests a more complex history involving 
both external minor accretion events and internal mechanisms that could influence the evolution 
of the NGC4649 GC system (i.e., dynamical friction 	and/or disk shocking). Overall,
the emerging picture is a perturbed GC system, suggesting a still evolving NGC4649, perhaps caught 
in the moment of accreting some smaller satellite galaxy. 

This interpretation is not straightforward because other supporting evidence is still
lacking other than the overall kinematics of PN, GCs and the stellar component of 
NGC4649~\citep{das2011,arnold2013}. We have searched for the kinematical signature of 
gravitational interactions and accretion of 
satellites undergoing in NGC4649, by comparing the GC over-densities with the radial velocities 
for the sample of 121 GCs discussed by~\cite{lee2008} with spectroscopic observations. We do 
not find any obvious correlation. Moreover,~\cite{pierce2006} obtained spectra of 38 GCs in 
NGC4649 (all included in our sample) and did not find evidence that red GCs are significantly 
younger than blue GCs. However, the GC samples of~\cite{lee2008} and~\cite{pierce2006} are both 
sparse and do not provide the optimal coverage
of the over-density feature. 

The similarity of the radial density profiles of red and blue GCs in NGC4649 (per 
Figure~\ref{fig:radialprofiles}, left) would be in principle consistent with NGC4649 being the result 
of a major dry merger~\citep[see][]{arnold2013,shin2009}. However, our results 
suggest that the flattening of the radial density profiles of both color classes, instead than 
being the result of a global process with no azimuthal dependence, could be driven by the existence of
major features in the spatial distribution of GCs. In the case of NGC4649, the arc extending from 
the center of the galaxy to the 
$D_{25}$ isophote and more significant for red GCs than for blue GCs, could be the responsible 
for the overall similarity of the radial density profiles of the two GC color classes.

\subsection{The LMXB population of NGC4649}
\label{subsec:lmxbngc4649}

It has been recognized since their early discovery in the Milky Way that dynamical formation of 
LMXBs in GCs is highly efficient~\citep{clark1975}; some of these binaries could then be either kicked 
out from the parent cluster or be left in the stellar field after cluster 
disruption~\citep{grindlay1984}~\citep[see also][for early-type galaxies]{kundu2002}. Therefore at least 
some field LMXBs could have been formed in GCs. However, LMXBs can also evolve 
from native binary systems in the stellar field~\citep[see review from][]{verbunt1995}. The effectiveness 
of either or both formation channels 
for the LMXB populations of both the Milky Way and external galaxies is still a matter of debate. Based on 
statistical considerations on the properties of the LMXB populations detected with Chandra in early-type 
galaxies, there is some indication that field evolution is a viable formation 
channel~\citep{juett2005,irwin2005,kim2009}, and that the 
LMXBs detected in the stellar field may be of mixed origin~\citep{kim2009,mineo2013}.

As is the case for GCs, LMXBs are also objects that can be detected individually in elliptical galaxies 
with the resolution of Chandra~\citep[see][]{fabbiano2006}. Therefore, their spatial distribution may provide some 
constraints on their origin. In particular, native field LMXBs should - within 
statistics - trace the stellar surface brightness distribution of their parent galaxy. 
Although there has been some debate on the radial profiles of LMXBs and their comparison 
with those of the stellar surface brightness and of GCs~\citep{kim2006,kundu2007}, in NGC4649 
there is clear evidence~\citep{mineo2013} that GC-LMXBs in red and blue GCs follow the same radial 
profiles as their parent GC populations, while field LMXBs in NGC4649 are radially distributed like the stellar 
surface brightness. However, within the limited sample statistics, LMXBs in red GCs are also distributed as 
the stellar light, with the exception of a marked lack of both red GCs and associated LMXBs in the 
centermost region. Therefore these results are still consistent with a mixed origin for the field LMXB population.

\begin{figure*}[]
	\includegraphics[height=8cm,width=8cm,angle=0]{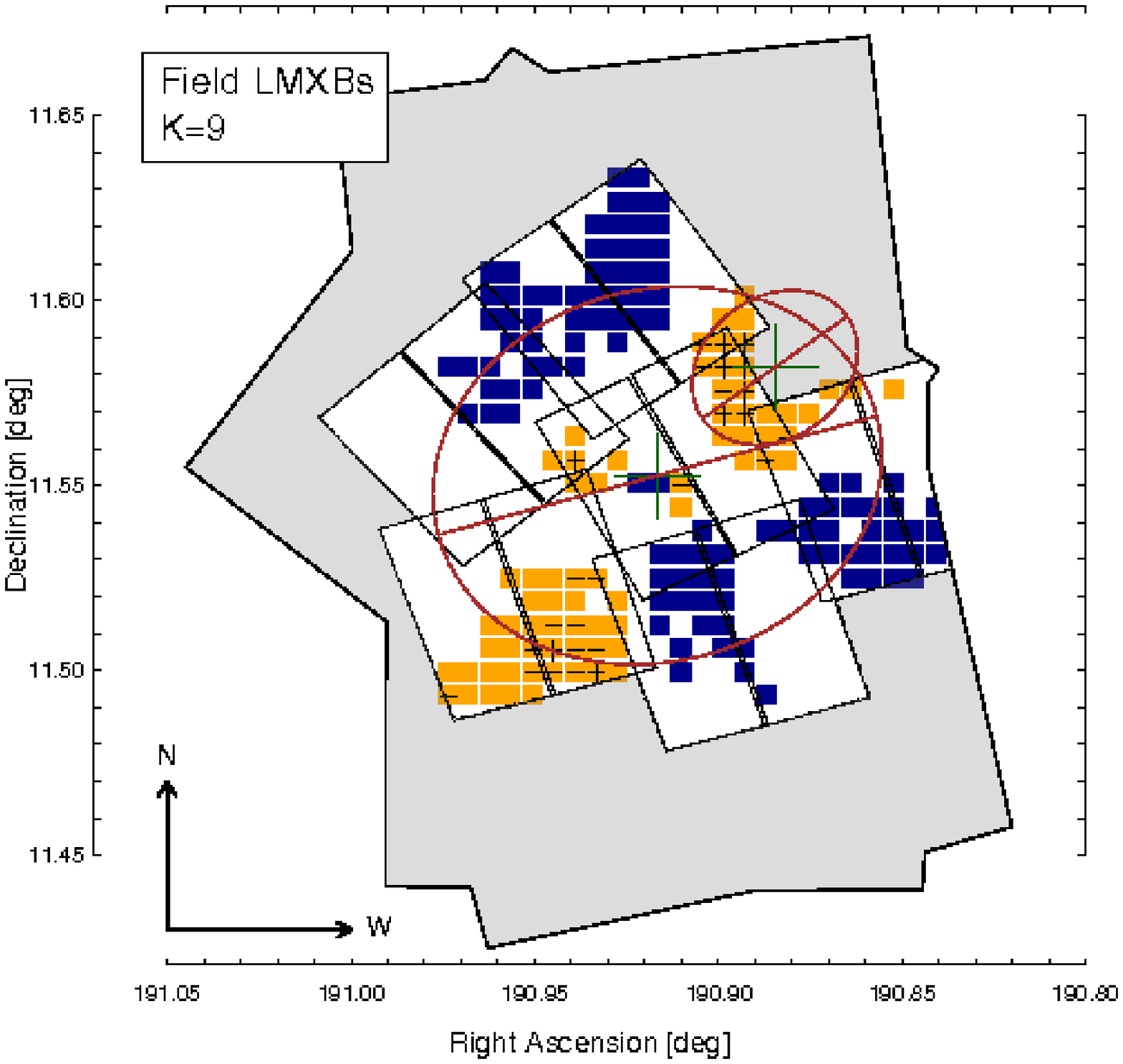}
	\includegraphics[height=8cm,width=8cm,angle=0]{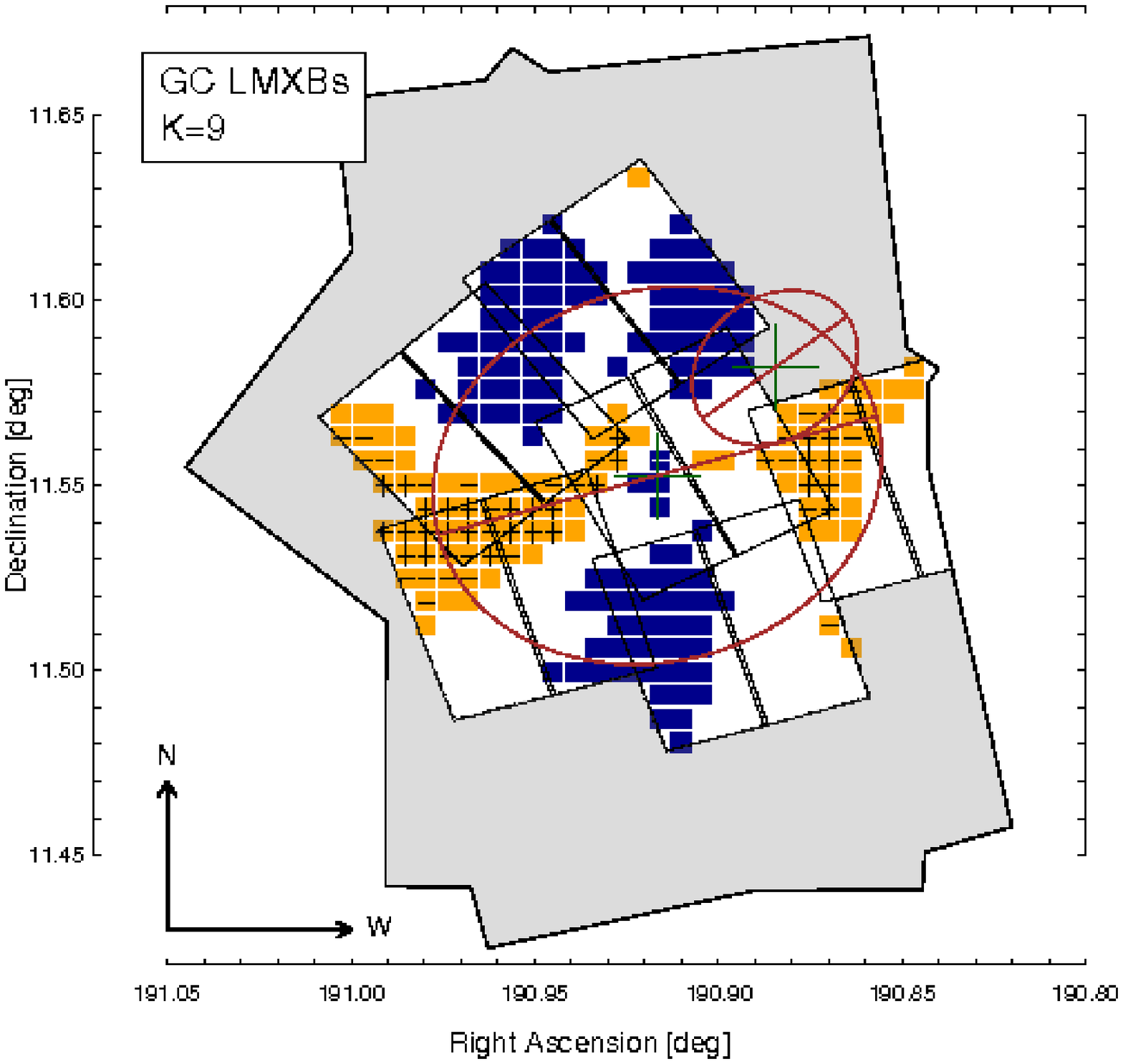}\\
	\includegraphics[height=8cm,width=8cm,angle=0]{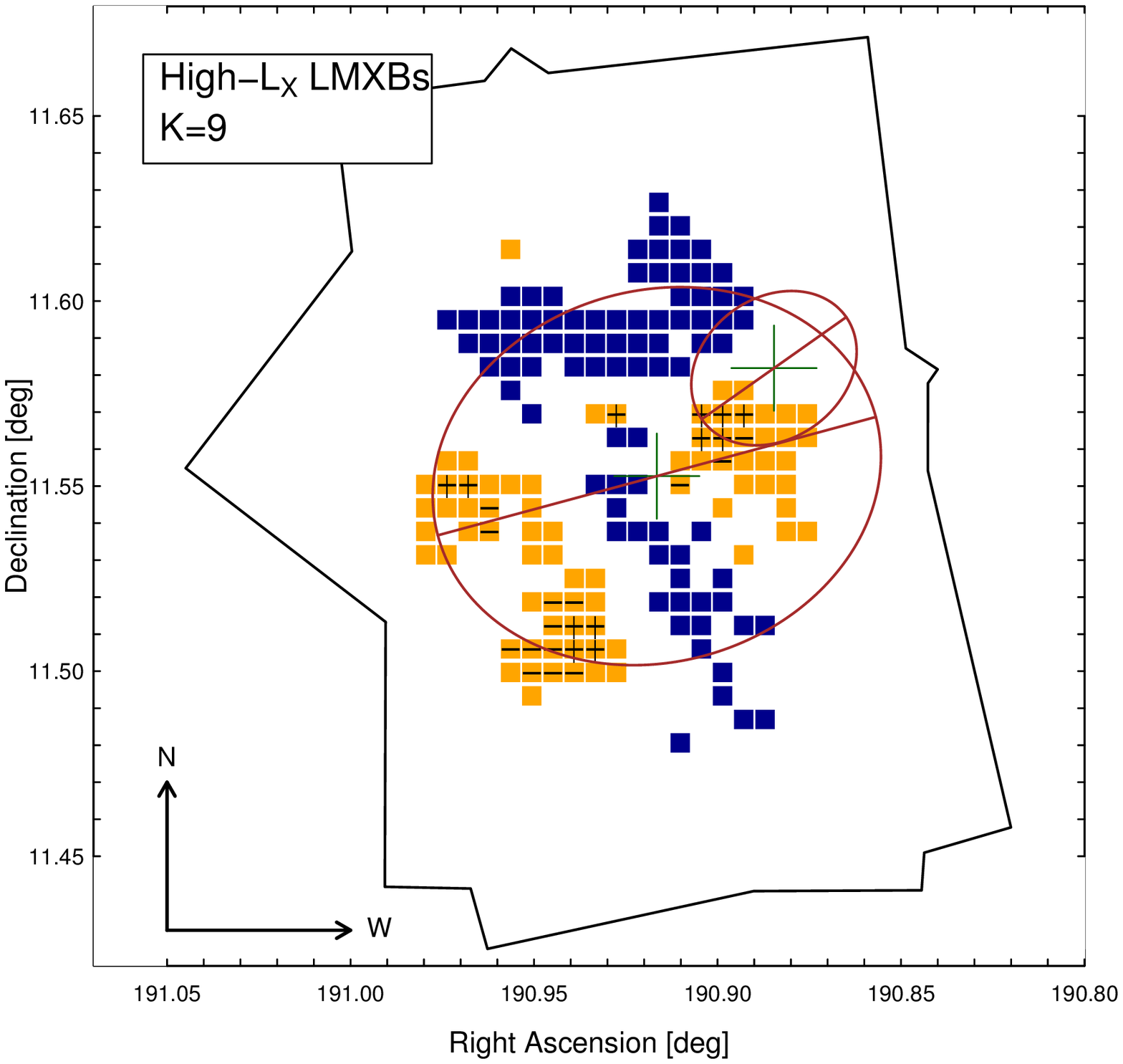}
	\includegraphics[height=8cm,width=8cm,angle=0]{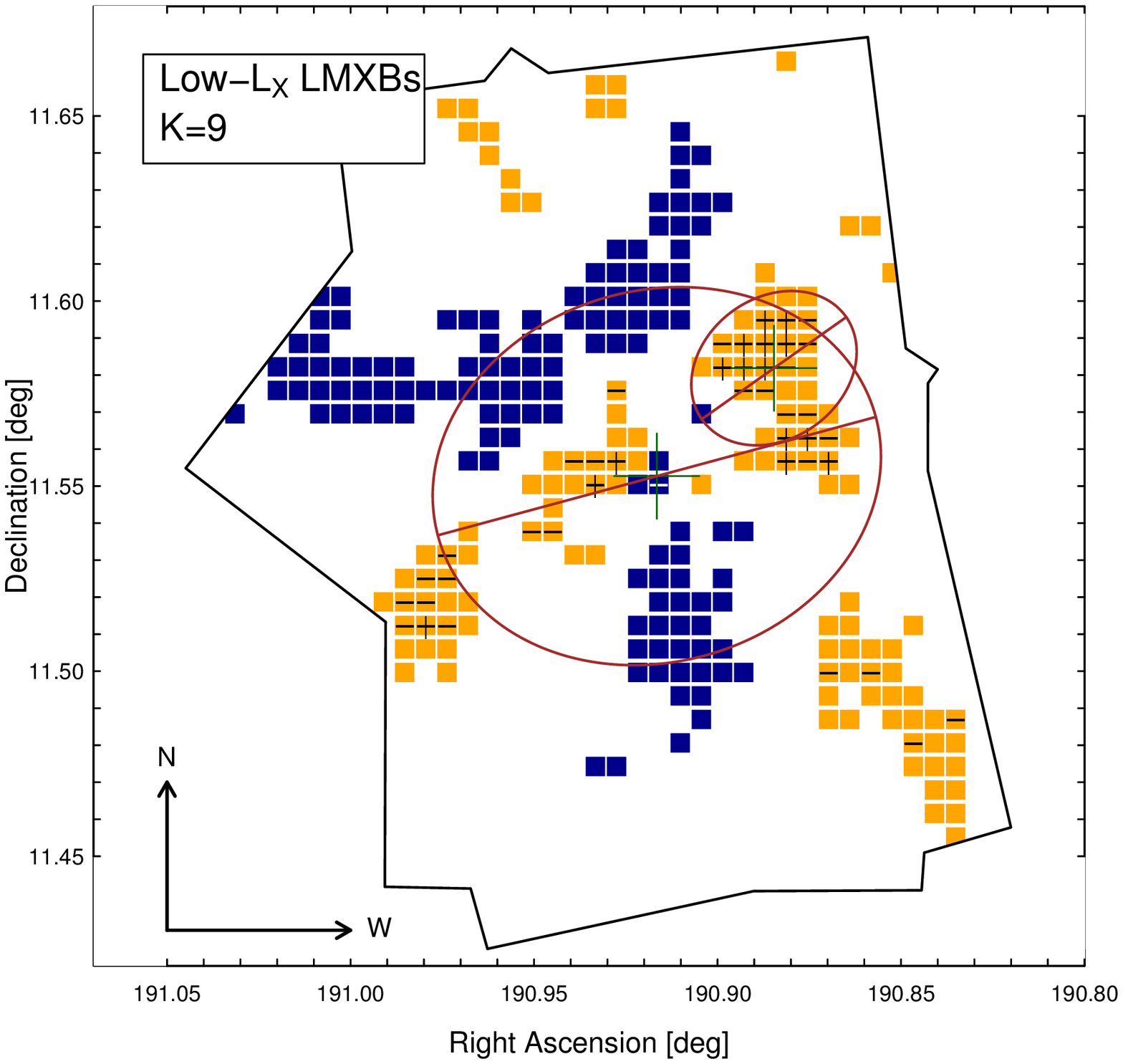}\\	
	\caption{Upper panels: positions of the $\!>\!1\sigma$ residuals for field (left) and GCs-LMXBs
	(right) samples of LMXBs for $K\!=\!9$. Positive (orange) pixels $\!>\!2\sigma$ are indicated 
	with a horizontal line; $\!>\!3\sigma$ with a cross. 
	Lower panels: positions of the $\!>\!1\sigma$ residuals for low- (left) and high-luminosity (right) 
	LMXBs samples for $K\!=\!9$. Positive (orange) pixels $\!>\!2\sigma$ are indicated 
	with a horizontal line; $\!>\!3\sigma$ with a cross. In all plots the $D_{25}$ elliptical isophotes 
	of both galaxies and the footprints of the HST and {\it Chandra} observations used to 
	extract the GCs and LMXBs catalogs respectively are shown for reference.}
	\label{fig:sigma_pixels_maps_lmxb}
\end{figure*}

We have found (Section~\ref{subsec:lmxbdensity}) that the 2D distribution of LMXBs shows a positive 
excess over a simple radial distribution with no azimuthal dependences, in the region near to the eastern 
major axis of NGC4649, where the excess feature of GCs is also observed 
(see Figure~\ref{fig:sigma_pixels_maps_lmxb}, which displays the higher significance individual
pixels for the residual distribution 
of different classes of LMXBs). Not surprisingly, the excess residuals of the distribution of LMXBs in 
GC, which are overwhelmingly associated with red GCs~\citep{luo2013,mineo2013}, resemble 
those of the red GCs. However, the 2D distribution of field LMXBs also shows large-scale anisotropy, 
which would not be expected if these sources were solely associated with native field binaries. The 
excess residuals appear to be adjacent to, but not overlapping, those seen in the 2D distribution of 
GC-LMXBs. This excess of field LMXBs is dominated by high 
luminosity LMXBs ($L_{X}\!>\!10^{38} \mathrm{erg\ s}^{-1}$). A second excess clump of these 
high-luminosity LMXBs is instead associated with the eastern end of the major axis; this over-density of
high-luminosity LMXBs is composed by equal number of GC-LMXBs and field LMXBs
(see right plot in Figure~\ref{fig:positions}). Positive residuals in the low luminosity LMXB 2D distribution 
are found extending along the eastern major axis, mostly associated to the positions of field LMBXs (compare 
right plot in Figure~\ref{fig:positions}); given their spatial distribution, though, these are also 
likely to have been originated in GCs. 

Of the two over-densities visible in the residual map
generated with all the LMXBs (Figure~\ref{fig:2dmapsngc4649_lmxb}) in the N-E and S-W corners
of the {\it Chandra} field, only the S-W structure can be seen in the residual map generated by 
low-L$_{X}$ LMXBs; neither is see in the high-L$_{X}$ residuals (Figure~\ref{fig:sigma_pixels_maps_lmxb}). 
The differences in the significance of these two LMXB over-density 
structures are due to the higher values of the radial density profile of high-L LMXBs at larger radii 
compared to that of low-L at large radii (see Figure~\ref{fig:radialprofiles}), while
both over-densities are composed of similar fractions of high and low luminosity X-ray 
sources. The N-E structure contains a total of 23 X-ray sources (12 low-luminosity and 11 
high-luminosity), while the S-W region contains 45 total sources (27 low-luminosity and 18 high-luminosity).

The presence of this high-L excess of field sources, to the south of the GC (and GC-LMXB) excess feature, 
suggests that these sources cannot have been formed in GCs and ejected in the field, unless 
the streaming motion of the GCs and of the ejected LMXBs are subject to different accelerations.
The escape velocity for a typical GC can be 
as low as $\sim$30 km s$^{-1}$~\citep{morscher2013}, where the range of line of sight velocities observed 
for GCs in massive elliptical galaxies extends from few to few thousands km s$^{-1}$~\citep{strader2011}.
This suggests that lacking other forces, the LMXBs ejected should settle into a cloud surrounding their
parent GCs. However, dynamical friction may cause the observed positional offset between the 
over-density of red GCs observed along the E side of the major axis (upper right plot in 
Figure~\ref{fig:sigma_pixels_maps_lmxb})
and the field LMXBs structure observed along the $D_{25}$ isophote south of the main GCs over-density
(upper left plot in Figure~\ref{fig:sigma_pixels_maps_lmxb}). We assume a 
standard mass for the LMXBs-hosting GCs 
$M_{GC}\!=\!10^{7}M_{\sun}$ consistent with their average absolute 
magnitudes~\citep{strader2012}. We further assume a circular velocity $v_{circ}\!\sim\!450$ km s$^{-1}$ 
based on the results from~\cite{humphrey2008}. 
Using as starting radial distance half the length of the major axis of the galaxy equivalent to 
$r\!=\!15$ kpc, we estimate the time required to recreate the angular and 
radial separation of the field LMXBs and GCs LMXBs over-densities with equation (7-26) of~\cite{binney2008}, 
that evaluates the time required for a GC moving on a circular orbit of radius $r_{i}$ with velocity $v_{c}$ 
to reach the center of the galaxy ($\ln{\Lambda}\!=\!10$ is assumed, based on typical values of the GC 
physical parameters):

\begin{equation}
	t_{\mathrm{fric}}\!=\!\frac{1.17}{\ln{\Lambda}}\frac{r_{i}^{2}v_{c}}{GM}
\end{equation} 

We obtain a time scale of $\sim\!5$ Gyr. This timespan is shorter than the average GC age of $\sim\!$10 
Gyr estimated by~\cite{pierce2006}, and within the range of single GCs measurements by these 
authors. Accurate kinematic measurements for GC population in NGC4649 would be needed
to test this hypothesis. Also, we point out that our simple evaluation of the effects of the 
dynamical friction should only be considered indicative. A full set of dynamical simulations
of the evolution of the GC system of NGC4649 as the results of satellite accretion and 
mergers is required. 
Alternatively, one could speculate that
the merging event originating the GC over-density may have also resulted in a compressed gaseous 
stream from the disrupted satellite galaxy, with subsequent star and X-ray binary formation. The 
higher luminosity of the over-dense field X-ray sources would be consistent with a younger age than 
the overall LMXB population~(\citealt{fragos2013a,fragos2013b}).

\subsection{The X-ray binary population of NGC4647}
\label{subsec:lmxbngc4647}

NGC4647 is a spiral galaxy in Virgo that can be seen to the N-W of NGC4649. Based on its aspect
and photometric properties,~\cite{lanz2013} conclude that it is tidally interacting with NGC4649. 
There is a concentration of X-ray sources (positive residuals) associated with this spiral galaxy. 
The high-luminosity X-ray sources concentrate in the southern half of NGC4647. This is the part
of the galaxy showing a prominent asymmetric arm suggesting interaction with NGC4649, and
exhibiting more intense star formation~\citep{lanz2013,mineo2013}. These luminous sources are 
likely younger HMXBs, associated with the ongoing intense star formation~\citep[e.g.][]{mineo2012}. 
Our results agree with the overall picture of tidal interaction between NGC4647 and NGC4649. 
NGC4647 may well be the next accreted satellite. 

\section{Summary and Conclusions}
\label{sec:conclusions}

Our analysis of the 2D projected distribution of the GC and LMXB populations of the giant Virgo elliptical 
NGC4649 (M60) has led to the following results:
\begin{itemize}
	\item There are significant 2D asymmetries in the projected distributions of both GCs and LMXBs.
	\item Red GCs are more centrally concentrated than blue GCs, as usually observed in elliptical 
	galaxies (e.g.,~\citealt{brodie2006}; see also~\citealt{mineo2013} for NGC4649). The same is also true 
	for high luminosity GCs (which tend to be red; see~\citealt{fabbiano2006}).
	\item The 2D distribution of GCs shows strong positive residuals along the major axis of NGC4649, 
	especially on the E side, with an arc-like curvature towards the N, beginning at a galactocentric 
	radius of $\sim\!4$ kpc and extending out to $\sim\!15$ kpc. This large-scale feature is more prominent in 
	the distribution of red GCs, but can still be seen in the 2D distribution of blue GC, which is overall 
	noisier. High-luminosity GCs tend to concentrate at the S end of the ``arc'', while low-luminosity 
	GCs are found in the Northern portion.
	\item The 2D distribution of LMXBs also shows positive excess in the E major axis side of NGC4649. 
	\item The anisotropies of the distribution of LMXBs in GC, which are overwhelmingly associated with 
	red GCs, resembles that of red GCs. However, field LMXBs also show large-scale anisotropies, 
	with positive excess to the S of the red GC feature; this excess is composed of high luminosity 
	LMXB ($L_{X}\!>\!10^{38} \mathrm{erg\ s}^{-1}$), which are also associated with the E end of 
	the major axis (these are partially associated to the position of GC-LMXBs). Positive excess in low 
	luminosity LMXB, mostly composed of field LMXBs, is found extending along the E major axis; 
	given their position, we speculate that these sources are likely to have formed in 
	GCs and subsequently ejected in the field.
	\item There is a definite concentration of LMXBs (positive residuals) associated with the spiral galaxy 
	NGC4647 in the N-W side of NGC4649. The higher X-ray luminosity portion of these sources tends 
	to concentrate in the southern half of NGC4647.
\end{itemize}

These results have important implications for the evolution of NGC4649. In particular, they suggest accretion
and merging of satellite galaxies, of which the GC anisotropic distribution is the fossil remnant. LMXB 
over-densities are found associated with those of red GCs, consistent with dynamic LMXB formation in 
these clusters. Surprisingly, there is also an over-density in the distribution of field LMXBs, to the south 
of the red GC over-density. This over density suggests that these LMXBs should be somewhat 
connected with the GC overdensity. The displacement, however, implies either (1) GC formation 
followed by ejection and dynamical friction drag on the parent GCs, or (2) field formation stimulated 
by compression of the ISM during a merging event. 

We also clearly detect the over-density of X-ray sources connected with the X-Ray Binaries (XRB) 
population of the 
companion spiral galaxy NGC4647. We note that the XRBs sources located in the southern half of 
the NGC4647, where the star formation
rate appears to be more intense (see~\cite{mineo2013}), are significantly more luminous that the X-ray 
sources observed in northern region of NGC4647, with $L_{X}\!>\!10^{38} \mathrm{erg\ s}^{-1}$. This result 
may suggest a younger XRB population. We speculate that this asymmetry may indicate the beginning 
of the tidal interaction of NGC4647 with NGC4649. 

Our analysis and the kinematics of NGC4649~\citep{das2011,coccato2013,arnold2013} 
suggest a complex assembly history for this galaxy, including one major merger and a sequence of satellite 
accretion events, with possibly continuous tidal stripping of GCs and PNs from nearby low-luminosity galaxies.
Explaining the origin of the over-density structures in the spatial distribution of GCs and of the luminosity and
color segregations within such features in NGC4649 will require both detailed numerical simulations 
with a realistic model of the galaxy potential and extensive campaign for the spectroscopic observation
required to reconstruct accurately the kinematics of the whole GC population. The detailed characterization 
of the 2D spatial distribution of GCs and LMXBs in NGC4649 that we have achieved will provide a new 
benchmark for future work aimed at the understanding of the dynamical evolution of this interesting
system.

\acknowledgements
We thank Jean Brodie for comments that have helped to improve the paper. 
%R. D'Abrusco gratefully acknowledges the financial 
%support of the US Virtual Astronomical Observatory, which is sponsored by the
%National Science Foundation and the National Aeronautics and Space Administration.
GF thanks the Aspen Center for Physics for hospitality (NSF grant 1066293).
TF acknowledges support from the CfA and the ITC prize fellowship programs. SM 
acknowledges funding from the STFC grant ST/K000861/1. This work 
was partially supported by the Chandra GO grant GO1-12110X and the associated HST 
grant GO-12369.01A, and the {\it Chandra} X-ray Center (CXC), which is operated by the 
Smithsonian Astrophysical Observatory (SAO) under NASA contract NAS8-03060.

{}

\end{document}